\documentclass[journal]{IEEEtran}
\usepackage{amsthm}
\usepackage{amsfonts}
\usepackage{amssymb}
\ifCLASSINFOpdf
  \usepackage[pdftex]{graphicx}
  \usepackage{graphicx}
  \graphicspath{{../pdf/}{../jpeg/}{../png/}}
  \DeclareGraphicsExtensions{.pdf,.jpeg,.png}
\else
  \usepackage[dvips]{graphicx}
  \graphicspath{{../eps/}}
  \DeclareGraphicsExtensions{.eps}
\fi
\ifCLASSOPTIONcompsoc
  \usepackage[caption=false,font=normalsize,labelfont=sf,textfont=sf]{subfig}
\else
\usepackage[caption=false,font=footnotesize]{subfig}
\fi
\usepackage{multirow}
\usepackage{epstopdf}
\usepackage{verbatim} 
\usepackage{xcolor}

\usepackage[english]{babel}
\usepackage{float}
\usepackage{xcolor}

\usepackage[hidelinks]{hyperref}
\usepackage{cite}
\ifCLASSINFOpdf
   \usepackage[pdftex]{graphicx}
   \else
\usepackage[dvips]{graphicx}
\fi
\usepackage[cmex10]{amsmath}
\usepackage[T1]{fontenc}
\newcommand*{\affmark}[1][*]{\textsuperscript{\dag}}
\usepackage{algorithm}
\usepackage[noend]{algpseudocode}
\makeatletter
\def\BState{\State\hskip-\ALG@thistlm}
\makeatother

\begin{document}
\title{Angularly Sparse Channel Estimation in Dual-Wideband Tera-Hertz (THz) Hybrid MIMO Systems Relying on Bayesian Learning}
\author{
Abhisha~Garg~\IEEEmembership{Graduate Student Member,~IEEE,} Suraj~Srivastava,~\IEEEmembership{Member,~IEEE,} Nimish~Yadav, Aditya~K.~Jagannatham,~\IEEEmembership{Senior Member,~IEEE} and Lajos Hanzo,~\IEEEmembership{Life Fellow,~IEEE}
\thanks{L. Hanzo would like to acknowledge the financial support of the Engineering and Physical Sciences Research Council projects EP/W016605/1, EP/X01228X/1 and EP/Y026721/1 as well as of the European Research Council's Advanced Fellow Grant QuantCom (Grant No. 789028).\\The work of Aditya K. Jagannatham was supported in part by the Qualcomm Innovation Fellowship; in part by the Qualcomm 6G UR Gift; and in part by the Arun Kumar Chair Professorship.\\The work is partially supported by the IEEE SPS Scholarship grant for $2023$.\\ Abhisha Garg, Nimish Yadav and Aditya K. Jagannatham are with the Department of Electrical Engineering, Indian Institute of Technology Kanpur, Kanpur-$208016$, India (e-mail: abhisha20@iitk.ac.in; nimishy20@iitk.ac.in; adityaj@iitk.ac.in). Suraj Srivastava is with the Department of Electrical Engineering, Indian Institute of Technology Jodhpur, Jodhpur, Rajasthan $342030$, India (email: surajsri@iitj.ac.in). L. Hanzo is with the School of Electronics and Computer Science, University of Southampton, Southampton SO17 1BJ, U.K. (email: lh@ecs.soton.ac.uk)}}
\maketitle
\begin{abstract}
Bayesian learning aided massive antenna array based THz MIMO systems are designed for \textit{spatial-wideband} and \textit{frequency-wideband} scenarios, collectively termed as the \textit{dual-wideband} channels. Essentially, numerous antenna modules of the THz system result in a significant delay in the transmission/ reception of signals in the time-domain across the antennas, which leads to spatial-selectivity. As a further phenomenon, the wide bandwidth of THz communication results in substantial variation of the effective angle of arrival/ departure (AoA/ AoD) with respect to the subcarrier frequency. This is termed as the \textit{beam squint effect}, which renders the channel state information (CSI) estimation challenging in such systems. To address this problem, initially, a pilot-aided (PA) Bayesian learning (PA-BL) framework is derived for the estimation of the Terahertz (THz) MIMO channel that relies exclusively on the pilot beams transmitted. Since the framework designed can successfully operate in an ill-posed model, it can verifiably lead to reduced pilot transmissions in comparison to conventional methodologies. The above paradigm is subsequently extended to additionally incorporate data symbols to derive a Data-Aided (DA) BL approach that performs joint data detection and CSI estimation. We will demonstrate that it is capable of improving the dual-wideband channel's estimate, despite further reducing the training overhead. The Bayesian Cram{\'e}r-Rao bounds (BCRLBs) are also obtained for explicitly characterizing the lower bounds on the mean squared error (MSE) of the PA-BL and DA-BL frameworks. Our simulation results show the improved normalized MSE (NMSE) and bit-error rate (BER) performance of the proposed estimation schemes and confirm that they approach their respective BCRLB benchmarks.
\end{abstract}
\begin{IEEEkeywords}
Terahertz, MIMO, dual-wideband, pilot-aided, data-aided, beam squint, Bayesian learning
\end{IEEEkeywords}
\IEEEpeerreviewmaketitle
\section{Introduction}
\vspace{-5pt}
Terahertz (THz) communication, relying on the (0.1-10) THz band, has the promise of achieving the ultra-high data rates targeted by next generation networks \cite{akyildiz2022terahertz}. When compared to the mmWave band that ranges from (30-100) GHz, the THz band can provide spectral windows with bandwidths in the range of 10 GHz or even higher \cite{shafie2022terahertz}. This can in turn lead to the development of rapid data transfer capabilities links with rates in the range of several hundreds of Gbps\cite{sarieddeen2020next}. However, it must be noted that the successful achievement of THz communication is enormously challenging, as the THz signal suffers from high atmospheric losses ranging from 5dB to 100dB \cite{yang2018atmospheric} and path losses ranging from 70dB to 140 dB \cite{shafie2022terahertz}, which can be attributed to the absorption by gases and the high carrier frequencies, respectively. In addition, THz signals are prone to blockage by physical barriers such as walls, human bodies etc \cite{sarieddeen2021overview}. Multiple-input multiple-output (MIMO) technology, which is capable of generating narrow directional beams having high array gains, is extremely well-suited for mitigating the losses occurring in the THz band. Fortuitously, numerous antenna elements can be embedded at both the transceiver ends in a THz system within a small form-factor \cite{srivastava2022hybrid}. At this juncture, it should be noted that the conventional transceiver architecture that necessitates a separate radio frequency (RF) chain per antenna can lead to a an excessive increase in the hardware complexity required to carry out the signal processing operations pertaining to modulation and demodulation \cite{faisal2020ultramassive}. A popular solution recommended in the latest research literature \cite{lin2018hybrid,li2020dynamic} to overcome this impediment is the novel hybrid signal processing architecture, wherein the signal processing operations are effectively split between the RF and baseband stages. This is in stark contrast to an orthodox transceiver design wherein the entire signal processing burden is borne by the baseband processor. The hybrid architecture coupled with large MIMO arrays can enormously increase the signal gain via optimal beamforming, while also limiting the hardware complexity through joint RF and baseband signal processing. 

Moreover, a typical wideband THz channel is frequency-selective in nature, due to the multipath induced delay spread, which is a characteristic feature of wireless signal propagation \cite{ning2021prospective}. This phenomenon is termed as frequency-wideband effect. A pair of crucial factors contributing to frequency selectivity in the THz band, even for LoS scenarios, are high atmospheric absorption and free-space path losses \cite{elayan2018terahertz}. The heightened atmospheric absorption primarily arises from water vapor molecules. Furthermore, due to frequency-selective nature of the absorption coefficient, the THz band is characterized by ultra-wideband transmission windows (TWs) and absorption coefficient peak regions (ACPRs) \cite{shafie2022terahertz}. Additionally, the presence of ACPRs in the THz band leads to pulse broadening in the time domain, presenting practical communication challenges within this frequency range. Owing to the short wavelength of the THz signal, a large antenna array leads to a progressively increasing phase shift in the received/ transmitted signals across the constituent antennas, corresponding to commensurate delays in the time domain \cite{dai2022delay}. This is termed as the spatial-wideband effect. This leads to the frequency dependent nature of the array response \cite{jian2019angle}. A general THz system that is exposed to both frequency-wideband and spatial-wideband effects is termed as a dual-wideband system, which is considered in our work. Furthermore, a practical wideband THz system suffers from the beam squint problem that can be attributed to the frequency dependent nature of the array response \cite{jian2019angle}. The availability of accurate channel state information (CSI), which plays a key role in beamforming, is of paramount importance in such systems. However, conventional techniques such as the least-squares (LS) and linear minimum mean squared error (LMMSE) estimators, require an excessive pilot overhead, owing to the large number of antennas, which reduces the spectral efficiency (SE). Thus, efficient CSI estimation techniques, which exploit the properties of the dual-wideband THz MIMO channel, hold the key for the successful realization of reliable high speed communication in these systems which also considers the beam squint effect. A concise overview of the relevant prior research is provided next.
\subsection{Review of the existing literature}
\vspace{-3pt}
Jornet and Akyildiz \cite{jornet2011channel}
developed a novel THz channel model accounting also for the molecular absorption and path loss, which was successfully leveraged to determine the capacity of THz systems. In \cite{saeed2020terahertz}, the principles of molecular absorption were combined with radioactive transfer theory to evaluate the path loss of THz signals for propagation distances in the range of a centimeter to a few meters, followed by the formulation of the THz channel model. Lin \textit{et al.}, \cite{lin2016terahertz}, developed a general array-of-subarrays architecture for a hybrid THz MIMO system, complemented by an innovative algorithm for adaptive distance-aware beamforming, which requires knowledge of the CSI. Therefore, it is important to note at this juncture that the end-to-end performance of a THz band hybrid MIMO system is crucially contingent on the availability of accurate CSI. This motivates the design of superior channel estimation techniques for THz MIMO applications, which can potentially maximize the performance. The well-known traditional channel estimation approaches, such as the LS and MMSE, perform poorly in the THz regime owing to their exclusive pilot overhead, which consequently leads to a low spectral efficiency \cite{zhang2020channel}. Nevertheless, a significant point to note is that the THz frequency range exhibits a strong directional propagation characteristic owing to using a large antenna array, hence resulting in a sparsely populated multipath channel in the angular domain. Naturally, sparse signal recovery techniques, which have gained popularity in recent research, can lead to verifiably improved CSI estimates in THz MIMO transceivers. In this context, the authors of \cite{srivastava2021data} design a superimposed pilot modulation technique, followed by sparse Bayesian learning (SBL)-based approaches for the estimation of the frequency-selective wideband channel. More specifically, they proposed a multiple-measurement vector (MMV)-based SBL algorithm for joint CSI estimation and data-detection. Venugopal \textit{et al.}, in their cutting-edge work \cite{venugopal2017channel}, proposed techniques for the estimation of a frequency-selective mmWave wideband channel in both single-carrier frequency domain equalization (SC-FDE) and orthogonal frequency-division multiplexing (OFDM)-based mmWave systems. In their work, the CSI estimation model was formulated as a sparse recovery problem, which was subsequently solved via a time-frequency estimation algorithm. The authors of \cite{srivastava2021fast} developed a block least-mean squares (BLMS) sparse adaptive framework for estimating the channel in a mmWave MIMO SC-FDE system. As a further advance, Gao \textit{et al.}, in their pioneering treatise \cite{gao2019wideband}, proposed a sparse beamspace CSI estimation technique for the estimation of a wideband mmWave MIMO system having lens antenna arrays. The novel scheme proposed in their work is based on successive support detection (SSD), which is inspired by the successive interference cancellation (SIC) principle of multi-user systems. Wang \textit{et al.} \cite{wang2018spatial} conceived new CSI estimation techniques for both frequency division duplex (FDD) as well as time division duplex (TDD) mmWave massive MIMO systems. Their framework also considers the spatial-wideband and frequency-wideband effects inherent in such systems and exploits the sparsity both in the angular and delay domains, while comprehensively covering both the uplink as well as the downlink. An attractive sensing-based high-resolution technique was designed in \cite{wang2019beam} that can successfully extract the physical parameters of the uplink CSI, viz., AoAs, time delays and complex channel gains, while accounting for the beam squint effect in mmWave hybrid MIMO systems. This is achieved by adaptively updating the dictionary matrix. A few of the latest treatises have also considered the impact of the spatial-wideband and frequency-wideband effects in massive MIMO THz channel estimation, which is of considerable practical importance. Dovelos \textit{et al.} \cite{dovelos2021channel} model an OFDM-based wideband THz system using a uniform planar array (UPA) under the spatial-frequency wideband effect. A sparse beamspace representation of the wideband THz channel is formulated in their work, followed by the application of the orthogonal matching pursuit (OMP) and the generalized simultaneous OMP (GSOMP) algorithm popular for the acquisition of the sparse channel. Chou \textit{et al.}, \cite{chou2022dual} addressed the spatial-wideband and frequency-wideband effect with respect to the estimation of a time-varying MIMO-OFDM channel in sub-Terahertz massive MIMO systems. The authors in their cutting-edge work developed a multiple measurement vector least squares compressed sensing (MMV-LS-CS) based CSI estimation technique, which utilizes the MMV-LS to recover the channel employing the previous channel support and the MMV-CS to detect any time-varying components in the residual signal.
\begin{table*}
    \centering
\caption{\small Boldly contrasting our contribution to the literature} \label{tab:lit_rev}
\begin{tabular}{|l|c|c|c|c|c|c|c|c|c|c|c|c|c|c|c|c|c|}    \hline

\textbf{Features} &\cite{srivastava2022hybrid} & \cite{lin2018hybrid} &\cite{li2020dynamic} & \cite{jian2019angle}& \cite{zhang2020channel} & \cite{wang2018spatial} &\cite{dovelos2021channel}  & \cite{chou2022dual}&\cite{rodriguez2018frequency} &\textbf{This paper} \\ 

 \hline

  THz hybrid MIMO

& \checkmark &  &  &  & &  & \checkmark & \checkmark & &  \checkmark\\

 \hline

 Dual-Wideband effect 

&  &  & \checkmark &  & & \checkmark &  & \checkmark &  & \checkmark \\

 \hline
 
Reflection losses

& \checkmark &  &  &  & &   & \checkmark & \checkmark &  & \checkmark\\

 \hline

Molecular absorption losses

& \checkmark &  &  &  & &  & \checkmark  & \checkmark &  & \checkmark\\

 \hline

Diffused-ray modeling

& \checkmark& \checkmark &  &  & &  &  &  &  & \checkmark\\

 \hline

BCRLB

& \checkmark & \checkmark &  & \checkmark & &  & \checkmark &  & \checkmark & \checkmark\\
\hline

SC-FDE system

&  &  & &  & &  &    &  &  & \checkmark\\
\hline
Data-Aided CSI acquisition

&  &  &  &  &  &  & &  &  &\checkmark\\
 \hline

\end{tabular}

\end{table*}
Although their pioneering work incorporates time-selectivity in sub-THz systems, no joint channel estimation and data detection is considered, which can lead to a remarkable enhancement in performance. To fill this knowledge gap, we conceive a novel DA Bayesian learning framework for reliable sparse CSI estimation and data-detection in a THz MIMO system, while considering also the dual-wideband effect. In Table I we boldly contrast our contributions to the literature. The next section briefly describes the main contributions of our work in more details.
\vspace{-6pt}
\subsection{Contributions of this work}
\vspace{-2pt}
\begin{enumerate}
    \item A hybrid frequency-dependent THz MIMO channel model is developed by considering the dual-wideband effects, which additionally incorporates the absorption, reflection as well as free space losses for a uniform linear array (ULA) at the base station (BS) serving a single user. Notably, our THz channel model incorporates the spatial wideband effect to construct a frequency-dependent model of the array response, while also considering the frequency-wideband effect, which accounts for multipath propagation. Existing literature on THz channel estimation, such as \cite{dovelos2021channel} and \cite{chou2022dual}, have solely focused on the spatial-wideband effect and did not include the frequency-wideband effect by considering multipath delays and the sampled version of transmit pulse shaping filter response. Therefore, the novel contribution of our work lies in considering the dual wideband effects, which are missing in the existing literature such as \cite{sha2021channel}, \cite{elbir2022federated}. Moreover, an efficient zero-padded (ZP) block structure is proposed to transform the time domain (TD) signal processing problem into an equivalent frequency domain (FD) one, which remarkably reduces the complexity of THz MIMO CSI estimation as well as signal detection. Additionally, this novel transmission structure requires very few pilot beams to excite the numerous angular modes of the channel.
    \item A sparse signal recovery problem is formulated for the THz MIMO channel. A pair novel pilot-aided Bayesian learning (PA-BL) and data-aided Bayesian learning (DA-BL)-based methodologies are proposed for estimating the sparse beamspace dual-wideband channel. Since it can operate in ill-posed scenerios, it requires significantly fewer pilot transmissions in comparison to the traditional LS and MMSE schemes.
    \item The Bayesian Cram{\'e}r-Rao lower bounds (BCRLB) are derived for the mean-square error (MSE) of both the PA and DA estimators proposed in the THz regime. These serve as excellent benchmarks to gauge the performance of the proposed algorithm.
    \item Simulation results are presented for quantifying the CSI estimation and data-detection performance achieved in the dual-wideband THz MIMO system considered. It is shown that the normalized MSEs (NMSEs) of the PA-BL and DA-BL algorithms approach their respective BCRLBs, which demonstrates their efficiency. This is especially significant considering the fact that they do not require prior knowledge of either the statistics or of the support of the sparse CSI.
\end{enumerate}
\vspace{-8pt}
\subsection{Organization of the paper}
The paper is structured as follows. Section II derives the system model of a hybrid massive MIMO system in the THz regime utilizing the SC-FDE framework. Section III motivates and models the dual-wideband effect in the above system and also characterizes the loss induced by the beam squint effect. Section IV determines the model for sparse CSI estimation in the wideband THz system. Section V describes the PA-BL approach proposed for CSI estimation, with the requisite mathematical foundation. Next, Section VI chronicles the DA-BL approach, which yields an improved performance over the previous paradigm. Section VII outlines our simulation results, while Section VIII concludes the paper.
\vspace{-10pt}
\subsection{Notation}
The following notation is used throughout this paper. Letters in upper-case boldface such as $\mathbf{A}$ represent matrices, while letters in lower-case boldface denote vectors. Superscripts $(.)^{\mathit{T}}$, $(.)^{*}$, $(.)^{\textrm{-1}}$, $(.)^{\mathit{H}}$, and $(.)^{\dagger}$ represent the transpose, conjugate, inverse, Hermitian and pseudo-inverse operators, respectively. The quantity $\textrm{blkdiag}\left(\mathbf{A}_{\textrm{1}}, \mathbf{A}_{\textrm{2}},\cdots,\mathbf{A}_{\mathit{N}}\right)$ represents the block-diagonal matrix with matrices $\mathbf{A}_{\textrm{1}}, \mathbf{A}_{\textrm{2}},\cdots,\mathbf{A}_{\mathit{N}}$ on the principal diagonal, which are not necessarily square. The vectorization operation that stacks the columns of any matrix $\mathbf{A}$ is represented by $\textrm{vec}(\mathbf{A})$, while the inverse of the same operation is represented by $\textrm{vec}^{-1}(\mathbf{a})$. The operators $*$ and $\otimes$ represent the linear convolution and Kronecker product, respectively, while $\textrm{Tr}(.)$ represents the matrix trace operator. The following standard property of the $\textrm{vec}(.)$ operation, $\textrm{vec}(\mathbf{A}\mathbf{B}\mathbf{C}) = (\mathbf{C}^{T}\otimes\mathbf{A})\textrm{vec}(\mathbf{B})$, is also used in the paper. The operators $\parallel.\parallel_{\textrm{2}}$ and $\parallel.\parallel_{\mathit{F}}$ represent the $l_{\textrm{2}}$-norm and Frobenius norm, respectively, of vectors and matrices. The expectation operator is denoted by $\mathbb{E}\left\{.\right\}$, while the absolute value is denoted by $| \ . \ |$; the symmetric complex Gaussian distribution with mean $\boldsymbol{\mu}$ and variance $\boldsymbol{\Sigma}$ is represented by $\mathcal{CN}\left(\boldsymbol{\mu},\boldsymbol{\Sigma}\right)$. 
\vspace{-5pt}
\section{SC-FDE Based THz Hybrid MIMO System Model}
\begin{figure*}[t]
\centering
\includegraphics[scale=0.33]{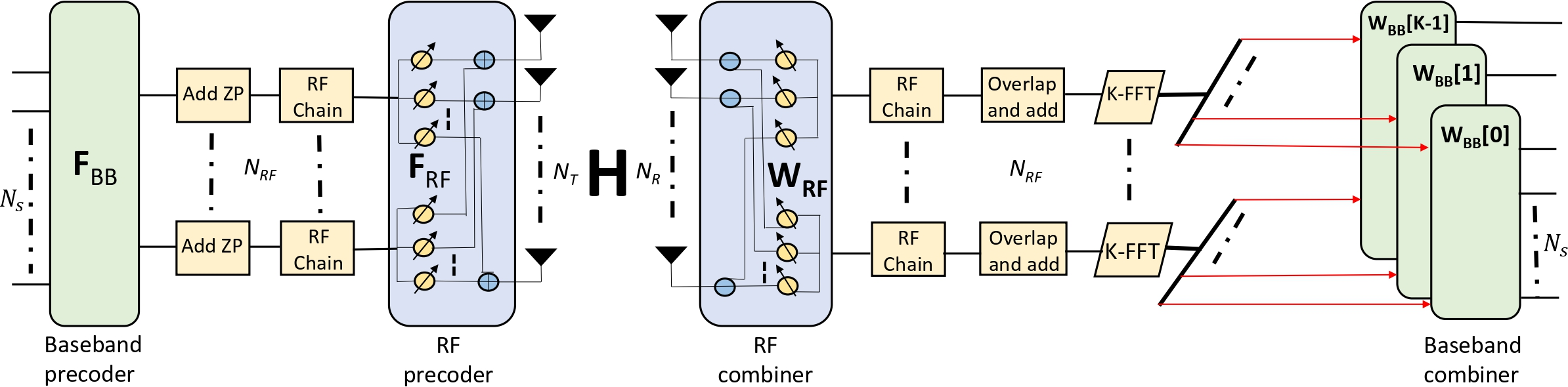}
\vspace{-4pt}
\caption{Schematic representation of SC-FDE-based wideband THz hybrid MIMO system}
\label{fig:THz MIMO}
\end{figure*}
Consider a single carrier wideband THz hybrid MIMO system that transmits $N_{s}$ data streams from a transmitter equipped with $N_{T}$ transmit antennas (TAs), to a receiver having $N_{R}$ receive antennas (RAs). There are $N_{RF}$ RF chains at the transmitter and the receiver, where $N_{s}\leq N_{RF}\ll \mathrm{min}(N_{T},N_{R})$. The THz hybrid MIMO transmitter is equipped with a baseband precoder in cascade with an RF precoder, represented by the matrices $\mathbf{F_{\mathit{BB}}}$ $\in$ $\mathbb{C}^{N_{RF}\times N_{s}}$ and $\mathbf{F_{\mathit{RF}}}$ $\in$ $\mathbb{C}^{N_{T}\times N_{RF}}$, respectively. In a similar manner, at the RAs, the hybrid processing module is comprised of an RF beamformer $\mathbf{W_{\mathit{RF}}}$ $\in$ $\mathbb{C}^{N_{R}\times N_{RF}}$ in cascade with a baseband combiner $\mathbf{W_{\mathit{BB}}}$ $\in$ $\mathbb{C}^{N_{RF}\times N_{s}}$. The hybrid signal processing architecture of the THz MIMO system is shown in Fig. \ref{fig:THz MIMO}. Since the RF transmit precoder (TPC) matrix $\mathbf{F_{\mathit{RF}}}$ and RF receiver combiner (RC) matrix $\mathbf{W_{\mathit{RF}}}$ are constrained to exclusively comprise digitally controlled phase-shifters, having elements obeying $|\mathbf{F_{\mathit{RF}}}(i,j)| = \frac{1}{\sqrt{N_{T}}}$, $|\mathbf{W_{\mathit{RF}}}(i,j)| = \frac{1}{\sqrt{N_{R}}}$ \cite{venugopal2017channel}. Furthermore, $\mathbf{F}_{RF}$ and $\mathbf{W}_{RF}$ can be represented as $\mathbf{F}_{\mathit{RF}} = \left[\mathbf{f}_1, \mathbf{f}_2,\cdots, \mathbf{f}_{N_{RF}}\right] $ and $\mathbf{W}_{\mathit{RF}} = \left[\mathbf{w}_1, \mathbf{w}_2,\cdots, \mathbf{w}_{N_{RF}}\right]$, respectively. Considering a frequency-selective wideband channel in the THz band comprised of $\mathit{L}$ delay taps, the $\mathit{l}$th delay tap can be expressed by the complex channel matrix $\mathbf{H}_{\mathit{l}}$, $0 \leqslant \mathit{l} \leqslant \mathit{L}-1$, of size $N_R \times N_T$. The symbol vector that is transmitted at time instant $\mathit{t}$ is denoted by ${\mathbf{s}}(\mathit{t})$ $\in$ $\mathbb{C}^{N_{s}\times 1}$. After baseband combining, the output signal vector at the $\mathit{t}$th time instant is given by
\begin{align}
\check{\mathbf{y}}({\mathit{t}}) = \mathbf{W}\mathit{_{BB}^{H}}\mathbf{W}\mathit{_{RF}^{H}}[\mathbf{H}_{\mathit{t}}\ast \mathbf{x}(\mathit{t})] + \mathbf{W_{\mathit{BB}}^{\mathit{H}}}\mathbf{W_{\mathit{RF}}^{\mathit{H}}}\mathbf{v({\mathit{t}})}, \label{time_dom_output}
\end{align}
where $\mathbf{{x}}(\mathit{t}) = \mathbf{F_{\mathit{RF}}}\mathbf{F_{\mathit{BB}}}{\mathbf{s}}(\mathit{t}) \in $ $\mathbb{C}^{N_{\mathit{T}}\times 1}$ represents the precoded symbol vector and $\mathbf{v({\mathit{t}})} \in \mathbb{C}^{N_{\mathit{R}}\times 1}$ represents the additive noise vector that follows the distribution $\mathcal{CN}(\mathbf{0}_{N_{R}\times 1}, \sigma_{\mathit{v}}^{2}\mathbf{I}_{N_{R}})$. The operator $\ast$ represents linear convolution. The equivalent FD model of this system is developed next.
\subsection{SC-FDE channel estimation model}
\begin{figure}
\centering
\includegraphics[scale=0.75]{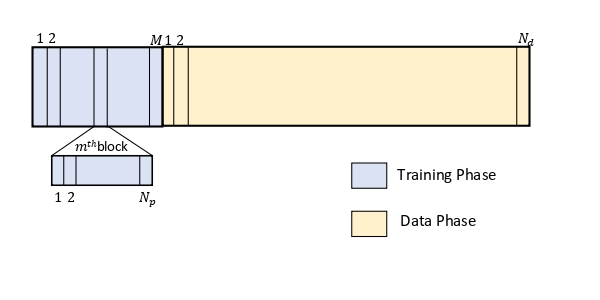}
\vspace{-28pt}
\caption{Frame structure employed in the wideband THz hybrid MIMO system using SC-FDE.}
\label{fig:frame}
\end{figure}

Again Fig. \ref{fig:THz MIMO} depicts the architecture of a hybrid THz MIMO system relying on FD signal processing. Each transmit frame duration in the system is divided into training and data phases, as shown in Fig. \ref{fig:frame}. The training phase is comprised of $\mathit{M}$ blocks with each individual block consisting of $N_{p}$ pilot vectors. Let $\mathbf{u}_{\mathit{m}}^{(\mathit{p})} \in \mathbb{C}^{\mathit{N_{RF}}\times 1}$ denote the $\mathit{p}$th complex pilot vector, which follows the relationship $0\leqslant \mathit{p}\leqslant \mathit{N_{p}-\mathrm{1}}$ for every $\mathit{m}$th block \cite{srivastava2021fast}. Prior to transmitting the pilot sequence on an individual RF chain, $\mathit{L} - 1$ zeros are appended in each $\mathit{m}$th block to generate a zero-padded (ZP) block of length $\mathit{K = N_{p} + L} - 1$. Similarly, zero-matrices of size $\mathit{N}_{R}\times \mathit{N}_{T}$ are appended to the MIMO channel taps $\mathbf{H}_l$ to obtain a block of length $\mathit{K}$. Thus, the zero-padded pilot vectors and THz MIMO channel taps can be described in a compact fashion as $\left\{ {\mathbf{u}_{\mathit{m}}^{(\mathit{q})}} \right\}_{\mathit{q}=0}^{\mathit{K}-1} = \left\{ {{\mathbf{u}_{\mathit{m}}^{(0)}},{\mathbf{u}_{\mathit{m}}^{(1)}},\cdots,{\mathbf{u}_{\mathit{m}}^{(\mathit{\mathit{N}_{p}}-1)}},\underbrace{\mathbf{0},\cdots,\mathbf{0}}_{\mathit{L}-1}} \right\}$ and $\left\{ \mathbf{H}_{\mathit{q}}\right\}_{\mathit{q}=0}^{\mathit{K}-1} = \left\{ \mathbf{H}_{0},\mathbf{H}_{1},\cdots,\mathbf{H}_{\mathit{L}-1},\underbrace{\mathbf{0},\cdots,\mathbf{0}}_{\mathit{N}_{p}-1} \right\}$, respectively. For each block $\mathit{m}$, the training RF TPC and RC are represented as $\mathbf{F}_{\mathit{RF,m}} \in \mathbb{C}^{\mathit{N_{T}}\times \mathit{N_{RF}}}$ and $\mathbf{W}_{\mathit{RF,m}} \in \mathbb{C}^{\mathit{N_{R}}\times \mathit{N_{RF}}}$, respectively. Thus, the output signal vectors $\widetilde{\mathbf{y}}_{\mathit{m}}(\mathit{q}) \in \mathbb{C}^{\mathit{N_{RF}}\times 1}$ corresponding to the $\mathit{m}$th block follow the model given by
\begin{align}
\widetilde{\mathbf{y}}_{\mathit{m}}(\mathit{q}) = \mathbf{W}_{\mathit{RF,m}}^{H}(\mathbf{H}_{q} \circledast_{\mathit{K}} \widetilde{\mathbf{u}}_{\mathit{m}}^{\mathit{(q)}}) + \mathbf{W}_{\mathit{RF,m}}^{H}\mathbf{v}_{\mathit{m}}(\mathit{q}).
\label{frame_output}
\end{align}
There $\widetilde{\mathbf{u}}_{\mathit{m}}^{\mathit{(q)}} = \mathbf{F}_{\mathit{RF,m}}{\mathbf{u}_{\mathit{m}}^{(\mathit{q})}}$, $\mathbf{v}_{\mathit{m}}(\mathit{q})$ denotes the complex white AWGN noises which follows the distribution $\mathcal{CN}(\mathbf{0}_{\mathit{N}_{r}\times 1},\sigma ^{2}_v\mathbf{I}_{\mathit{N}_{r}})$, and $\circledast_{\mathit{K}}$ is the circular convolution operation between sequences of length $K$. Consider the $\mathit{m}$th TD output block defined as $\left\{ \widetilde{\mathbf{y}}_{\mathit{m}}(\mathit{q})\right\}_{q=0}^{\mathit{K}-1} = \left\{ \widetilde{\mathbf{y}}_{\mathit{m}}(0), \widetilde{\mathbf{y}}_{\mathit{m}}(1),\cdots, \widetilde{\mathbf{y}}_{\mathit{m}}(\mathit{K}-1)\right\}$. At the output, employing the $\mathit{K}$-point fast fourier transform (FFT) operation, the FD representation that corresponds to the hybrid MIMO system in the THz band may be obtained as 
\begin{align}
\left\{\mathbf{y}_{\mathit{m}}[\mathit{k}] \right\}_{\mathit{k}=0}^{\mathit{K}-1} & = \textrm{FFT} (\left\{ \widetilde{\mathbf{y}}_{\mathit{m}}(\mathit{q})\right\}_{q=0}^{\mathit{K}-1}) \notag \\& =\left\{{{\mathbf{y}}_{\mathit{m}}[0],{\mathbf{y}}_{\mathit{m}}[1],\cdots,{\mathbf{y}}_{\mathit{m}}[\mathit{K}-1]}\right\}. \label{freq_output}
\end{align}
Thus, the received FD output vector ${\mathbf{y}}_{\mathit{m}}[\mathit{k}] \in\mathbb{C}^{\mathit{N}_{\mathit{RF}}\times 1}$ at the $\mathit{k}$th subcarrier can be expressed as:
\begin{align}
{\mathbf{y}}_{\mathit{m}}[\mathit{k}] = \mathbf{W}_{\mathit{RF,m}}^{\mathit{H}}\mathbf{H}[\mathit{k}]\mathbf{F}_{\mathit{RF,m}}\mathbf{u}_{\mathit{m}}[\mathit{k}] + \mathbf{W}_{\mathit{RF,m}}^{\mathit{H}}\check{\mathbf{v}}_{\mathit{m}}[\mathit{k}] ,\label{final_freq_output}
\end{align}
where $\mathbf{H}[k]$ depicts the THz MIMO channel as frequency response (CFR), whose model is detailed in the next section. The vector $\mathbf{u}_{\mathit{m}}[\mathit{k}] \in \mathbb{C}^{\mathit{N}_{RF}\times 1}$ denotes the $\mathit{k}$th FFT point of the pilot sequence $\mathbf{u}_{\mathit{m}}^{(\mathit{q})}$, and $\check{\mathbf{v}}_{\mathit{m}}[\mathit{k}] \in \mathbb{C}^{\mathit{N}_{R}\times 1}$ represents the $\mathit{k}$th FFT point of the noise sequence $\mathbf{v}_{\mathit{m}}^{(\mathit{q})}$. Upon exploiting the property of vec(.) operator as described in Section-I-C, \eqref{final_freq_output} can be recast as
\begin{align}
\mathbf{{y}}_\mathit{m}[\mathit{k}] = \underbrace{(\mathbf{u}_{\mathit{m}}^{\mathit{T}}[\mathit{k}]\mathbf{F}_{\mathit{RF,m}}^{\mathit{T}}\otimes \mathbf{W}_{\mathit{RF,m}}^{\mathit{H}})}_{\mathbf{\Phi}_{\mathit{m}}[\mathit{k}]\in\mathbb{C}^{\mathit{N}_{\mathit{RF}}\times \mathit{N}_{\mathit{R}}\mathit{N}_{\mathit{T}}}}\mathbf{h}[\mathit{k}] + \mathbf{{v}}_\mathit{m}[\mathit{k}], \label{beamspace_output}
\end{align}
where $\mathbf{h}[\mathit{k}] = \textrm{vec}(\mathbf{H}[\mathit{k}]) \in \mathbb{C}^{\mathit{N}_{T}\mathit{N}_{R}\times 1}$ represents the vectorized form of the THz MIMO channel model at the $k$th subcarrier, $\mathbf{\Phi}_{\mathit{m}}[\mathit{k}]$ represents the sensing matrix for the $\mathit{m}$th block of the $\mathit{k}$th subcarrier, while $\mathbf{{v}}_\mathit{m}[\mathit{k}] = \mathbf{W}_{\mathit{RF,m}}^{\mathit{H}}\check{\mathbf{v}}_{\mathit{m}}[\mathit{k}] \in \mathbb{C}^{\mathit{N}_{RF}\times 1}$ is the noise vector at the output of the RF combiner of the $\mathit{k}$th subcarrier. Note also that,
\begin{align}
\mathbb{E}\left\{\mathbf{{v}}_\mathit{m}[\mathit{k}]\mathbf{{v}}_\mathit{m}^{H}[\mathit{k}]\right\} & = \mathbb{E}\left\{\mathbf{W}_{\mathit{RF,m}}^{\mathit{H}}\check{\mathbf{v}}_{\mathit{m}}[\mathit{k}]\left(\mathbf{W}_{\mathit{RF,m}}^{\mathit{H}}\check{\mathbf{v}}_{\mathit{m}}[\mathit{k}]\right)^{H}\right\} \notag \\ &=\sigma^{2}_vK\mathbf{W}_{\mathit{RF,m}}^{\mathit{H}}\mathbf{W}_{\mathit{RF,m}} = \mathbf{R}_{m} . \label{Rm_deri_mmse}
\end{align}
The next section describes the THz channel incorporating the dual-wideband effect. 
\section{THz Channel Model with dual-wideband effect}
We now develop an explicit model for the THz MIMO channel incorporating \textit{dual wideband} effect, comprising of the spatial and frequency wideband effects. We commence by describing the spatial wideband effect. 
\subsection{Spatial-wideband effect}
Let us assume that a signal arrives at a ULA having $N$ identical antennas, indexed as $\left\{1,2,\cdots,\mathit{N}\right\}$, at the azimuth angle $\theta$. The frequency-dependent array steering vector $\mathbf{a}(\theta,f)$ of this system is given by
\begin{align}
\mathbf{a}(\theta,\mathit{f}) = \frac{1}{\sqrt{\mathit{N}}}\left [1, e^{-j\frac{2\pi}{\lambda}\mathit{d}_c \cos\theta},\cdots,e^{-j(\mathit{N}-1)\frac{2\pi}{\lambda}\mathit{d}_c \cos\theta}  \right ]^{T}. \label{beamsquint_arrayresponse}
\end{align}
The quantity $\lambda = \frac{c}{f}$ represents the wavelength as a function of frequency and $d_c$ is the fixed inter-antenna spacing, that is typically set as $\mathit{d}_c=\frac{c}{2 \times f_c}$. In general, the transceiver is designed based on the central carrier frequency $f_c$, which is an approximation of the frequency over the dynamic frequency range. Therefore, the approximation $\mathit{d}_c \approx \frac{\lambda}{2}$ only holds for a narrowband channel, i.e., when the bandwidth \textit{B} obeys $\mathit{B} \ll \mathit{f}_{\mathit{c}}$. However, this condition does not hold for a wideband signal spanning a wide frequency range. Let $\triangle \varphi_{\mathit{n},k}$ denote the spatial phase offset for the $\mathit{n}$th antenna with the AoA/AoD equal to $\theta$ at subcarrier $k$. Let $\mathit{f}_{\mathit{k}}$ represent the frequency of the $\mathit{k}$-th subcarrier that is defined as $\mathit{f}_{\mathit{k}} \triangleq \mathit{f}_{\mathit{c}} + \left(\mathit{k} - \frac{\mathit{K} + 1}{2}\right)\frac{\mathit{B}}{\mathit{K}}.$ From \eqref{beamsquint_arrayresponse}, the expression for $\triangle \varphi_{\mathit{n,k}}$ is given by
\begin{align}
    \triangle \varphi_{\mathit{n,k}} & = \frac{2\pi}{\lambda_k}(\mathit{n}-1)\mathit{d}_c \cos\theta = \frac{2 \pi}{\frac{c}{f_k}}(n-1)\frac{c}{f_c}\times\frac{1}{2} \cos \theta \notag \\ & = \pi(\mathit{n}-1) \delta_k \cos\theta, \label{ph_offset}
\end{align}
where $\delta_k = \frac{\mathit{f}_{k}}{\mathit{f}_{\mathit{c}}}$ represents the relative frequency of the subcarrier. Thus, the effective spatial AoA $\hat{\theta}(f_k)$ at subcarrier frequency $f_k$ is given by
\begin{align}
\hat{\theta}(\mathit{f_k}) = \textrm{arccos}(\delta_k \cos\theta). \label{spatial_angle}
\end{align}
Note that the spatial phase offset $\triangle \varphi_{n,k}$ varies by the subcarrier index as seen in Equation (\ref{ph_offset}), which implies that the effective spatial AoA/ AoD $\hat{\theta}(f_k)$ as defined in Equation (\ref{spatial_angle}) also varies with respect to the subcarrier index $k$. Thus, the spatial phase offset depends on the subcarrier frequency. Due to the large antenna aperture in the THz band, the propagation delay between the first and last antenna elements is non-negligible, hence the transmit symbol reaches different antennas with different delays. This is termed the \textit{spatial-wideband} effect. This spatial wideband effect is aggravated in the THz band due to having numerous antenna elements, coupled with the large bandwidth. By the same token, the variation in the frequency across the subcarriers leads to variations in the effective AoA/ AoD across the frequency band, which is known as the \textit{beam squint} effect. Thus, the beam squint effect is the manifestation of the spatial wideband effect in the frequency domain and therefore, the two phenomena are analogous \cite{wang2018spatial}, \cite{wang2019block}. Moreover, the magnitude of the beam squint is directly proportional to the extent of spatial wideband effect occurring in such systems, as discussed in Theorem-$1$ of \cite{wang2019beam}. Note that a majority of the literature ignores the above effect. Additionally, the THz channel experiences the frequency wideband effect as a consequence of the multipath delay spread. Therefore, a combination of spatial and frequency wideband effects in THz hybrid MIMO systems is termed the dual-wideband effect. From (\ref{spatial_angle}), one can further see that the effective spatial AoA/ AoD $\hat{\theta}(f_k)$ varies with respect to the subcarrier frequency, which depends on the relative frequency $\delta_k$. Hence, the normalized array gain $G(\theta, f_k)$ \cite{dovelos2021channel} for a wideband THz system for the $k$th subcarrier and angular direction $\theta$ is $G(\theta,f_k) = |\tilde{\mathbf{a}}^H(\theta,f_c)\tilde{\mathbf{a}}(\theta,f_k)|^2.$ The next section develops a model for the wideband THz channel suffering from the dual-wideband effect, which arises due to a combination of the spatial-wideband and frequency-wideband effects.
\subsection{THz channel model with dual wideband effect}
The array steering vector of the wideband THz channel exhibiting beam squint, characterized by frequency dependent angles, is given by
\begin{align}
\begin{split}
\tilde{\mathbf{a}}(\theta, \mathit{f_k}) = \frac{1}{\sqrt{N}}\left [1, e^{-j\pi\delta_k \cos(\theta)},\cdots, e^{{-j(\mathit{N}-1)\pi\delta_k \cos(\theta)}} \right ]^{T}. \label{beamsquint_spatial}
\end{split}
\end{align}
Thus, the dual wideband channel at the $\mathit{k}$th subcarrier, where $\mathit{k} = 1,\cdots,\mathit{K}$ can be modeled as
\begin{align}
\mathbf{H}[{\mathit{k}}] = \mathbf{H}^{\textrm{LoS}}[{\mathit{k}}] + \mathbf{H}^{\textrm{NLoS}}[{\mathit{k}}], \label{Hsub_Los_NLos}
\end{align}
where the LoS and NLoS components $\mathbf{H}^{\mathrm{LoS}}[k]$, $\mathbf{H}^{\mathrm{NLoS}}[k]$ are defined as
\vspace{-3pt}
\begin{align}
    \mathbf{H}^{\text{LoS}}[k] = \sqrt{N_T N_R} \alpha_{\text{L}}(f_k,d) \times \beta_{\tau_{\text{L}}} \times & A_T A_R \Tilde{\mathbf{a}}_R(\phi_{\text{L}}^R,f_k) \notag \\  &  \Tilde{\mathbf{a}}_T^H(\phi_{\text{L}}^T,f_k),
\end{align}
\vspace{-8pt}
\begin{align}
\mathbf{H}^{\textrm{NLoS}}[{\mathit{k}}]  = & {\sqrt{\frac{\mathit{N}_{\mathit{T}}\mathit{N}_{\mathit{R}}}{\mathit{N}_{\textrm{NLoS}}\mathit{N}_{\mathit{ray}}}}} \sum_{\mathit{z}=1}^{\mathit{N}_{\textrm{NLoS}}} \sum_{\mathit{j}=1}^{\mathit{N}_{\mathit{ray}}}\alpha_{\mathit{z},\mathit{j}}(\mathit{f_{\mathit{k}}},\mathit{d}) \times \beta_{\tau_{\mathit{z},\mathit{j}}} \times \notag \\  & \mathit{A}_{\mathit{T}}\mathit{A}_{\mathit{R}}
\tilde{\mathbf{a}}_{\mathit{R}}(\phi_{\mathit{z},\mathit{j}}^{\mathit{R}},\mathit{f_{\mathit{k}}})\tilde{\mathbf{a}}_{\mathit{T}}^{\mathit{H}}(\phi_{\mathit{z},\mathit{j}}^{\mathit{T}},\mathit{f_{\mathit{k}}}). \label{HNLos_sub}
\end{align}
The quantity $\beta_{\tau_{z,j}}$ is defined as $\beta_{\tau_{\mathit{z},\mathit{j}}} = \sum_{\mathit{l}=0}^{\mathit{K}-1}\mathit{p}(\mathit{l}\mathit{T}_{\mathit{s}}-\tau_{\mathit{z},\mathit{j}})e^{-j\frac{2\pi\mathit{k}\mathit{l}}{\mathit{K}}},  \forall\mathit{k},\mathit{l}$. The quantities $N_R$ and $N_T$ represent the number of RAs and TAs, $N_{ray}$ is the number of diffused rays from a scatterer, while $N_{\text{NLoS}}$ represents the number of clusters/ dominant scatterers present in the propagation environment. Note that, here a cluster represents a group of rays that arrive at the receiver around the same time and from the same direction \cite{ju2021millimeter}, which arise due to reflection and scattering from a dominant scatterer. The pulse shaping filter is represented by $p(.)$ and the sampling time by $T_s$, whereas $A_R$ and $A_T$ denote the RA and TA gains. The quantities $\tau_{(.)}$, $\phi^R_{(.)}$, $\phi^T_{(.)}$, $\alpha(.)$ represent the delay, AoA, AoD and complex path gain for the $l$th delay tap, the models for which are shown in the next section.
\subsubsection{Calculation of the complex-path gains}
The typical THz wireless channel incurs a high loss owing to various factors such as molecular absorption losses and molecular noise due to water vapour, among others. These aspects can be incorporated in the modeling of the complex-path gains $\alpha_L(f_k,d)$, $\alpha_{z,j}(f_k,d)$, which are a function of both the subcarrier frequency $f_k$ and transmission distance $d$. The complex path gain $\alpha_{(.)}(f_k,d)$ can be characterized by its magnitude and angle components as $\alpha_{(.)}(f_k,d) = |\alpha_{(.)}(f_k,d)|e^{j\omega}$. According to \cite{jornet2011channel}, the path-gain of the LoS component can be modeled as
\begin{align}
|\alpha_{\textrm{L}}(\mathit{f}_{\mathit{k}},\mathit{d})|^{2} = \mathit{L}_\textrm{spread}(\mathit{f}_{\mathit{k}},\mathit{d})\mathit{L}_\textrm{abs}(\mathit{f}_{\mathit{k}},\mathit{d}),
\label{los_path}
\end{align}
where $\mathit{L}_{\textrm{spread}}(\mathit{f}_{\mathit{k}},\mathit{d})$ represents the free-space loss, while $\mathit{L}_{\textrm{abs}}(\mathit{f}_\mathit{k},\mathit{d})$ represents the molecular absorption loss \cite{dovelos2021channel}, at frequency $f_k$, which are respectively given by
\vspace{-4pt}
\begin{align}
\mathit{L}_{\textrm{spread}}(\mathit{f}_{\mathit{k}},\mathit{d}) = \left (\frac{\mathit{c}}{4\pi\mathit{f}_{\mathit{k}}\mathit{d}}\right)^{2}, \mathit{L}_{\textrm{abs}}(\mathit{f}_{\mathit{k}},\mathit{d}) = e^{-\mathit{k}_{\textrm{abs}}(\mathit{f}_{\mathit{k}})\mathit{d}}. \label{Lspread}
\end{align}
The quantity $k_{\mathrm{abs}}(\mathit{f}_{\mathit{k}})$ represents the molecular absorption coefficient that is given $\mathit{k}_{\textrm{abs}}(\mathit{f}_{\mathit{k}}) = \sum_{\mathit{a},\mathit{g}}\mathit{k}_{\textrm{abs}}^{\mathit{a},\mathit{g}}(\mathit{f}_{\mathit{k}})$, where $\mathit{k}_{\textrm{abs}}^{\mathit{a},\mathit{g}}(\mathit{f}_{\mathit{k}})$ represents the measure of radiation absorption by the $a$th isotopologue of gas $g$. The parameter $\mathit{k}_{\textrm{abs}}^{\mathit{a},\mathit{g}}(\mathit{f}_{\mathit{k}})$ can be calculated from the HITRAN database given in \cite{jornet2011channel}. Similarly, for the complex path gain of the $\mathit{z}$th NLoS component, the magnitude of the $\mathit{j}$th diffuse-ray is mathematically given by
\vspace{-4pt}
\begin{align}
|\alpha_{\mathit{z},\mathit{j}}(\mathit{f}_{\mathit{k}},\mathit{d})|^{2} = \Gamma^{2}_{\mathit{z,\mathit{j}}}(\mathit{f}_{\mathit{k}})\mathit{L}_{\textrm{spread}}(\mathit{f}_{\mathit{k}},\mathit{d})\mathit{L}_\textrm{abs}(\mathit{f}_{\mathit{k}},\mathit{d}), \label{Nlos_path}
\end{align}
where $\Gamma_{\mathit{z,\mathit{j}}}(\mathit{f}_{\mathit{k}})$ represents the first-order reflection coefficient for the $\mathit{j}$th diffuse ray of the $z$th NLoS cluster \cite{srivastava2022hybrid}, which is defined as the product of the Fresnel Reflection Coefficient $\gamma_{z,j}(\mathit{f}_{\mathit{k}})$ and of the Rayleigh roughness factor $\varrho_{z,j}(\mathit{f}_{\mathit{k}})$ \cite{piesiewicz2007scattering}, that are respectively given by
\vspace{-6pt}
\begin{equation}
\Gamma_{z,j}(\mathit{f}_{\mathit{k}}) = \gamma_{z,j}(\mathit{f}_{\mathit{k}})\varrho_{z,j}(\mathit{f}_{\mathit{k}}), \notag
\end{equation}
\vspace{-8pt}
\begin{equation}
\gamma_{z,j}(\mathit{f}_{\mathit{k}}) = \frac{Z(\mathit{f}_{\mathit{k}})\cos(\theta_{\text{in}_{z,j}}) - Z_{0}\cos(\theta_{\text{ref}_{z,j}})}{Z(\mathit{f}_{\mathit{k}})\cos(\theta_{\text{in}_{z,j}}) + Z_{0}\cos(\theta_{\text{ref}_{z,j}})}, \notag
\end{equation}
\vspace{-8pt}
\begin{equation}
\varrho_{z,j}(\mathit{f}_{\mathit{k}}) = e^{-\frac{1}{2}\left(\frac{4\pi\mathit{f}\sigma \cos\left(\theta_{\text{in}_{z,j}}\right)}{\mathit{c}}\right)^{2}},\label{fres_ref_eqn}
\end{equation}
where $\theta_{\text{in}_{z,j}}$ denotes the angle of incidence, $\theta_{\text{ref}_{z,j}} = \sin^{-1}\left(\sin(\theta_{\text{in}_{z,j}})\frac{\mathit{Z}(\mathit{f}_{\mathit{k}})}{\mathit{Z}_{0}}\right)$ represents the angle of refraction, $\mathit{Z}_{0} = 377\Omega$ refers to the intrinsic impedance of free space and $Z(\mathit{f}_{\mathit{k}})$ represents the characteristic impedance of the reflecting medium given \cite{piesiewicz2007scattering} by $Z(\mathit{f}_{\mathit{k}}) = \sqrt{\frac{\mu_{0}}{\varepsilon_{0}\left(\mathit{n}^{2} - (\frac{\varsigma\mathit{c}}{4\pi\mathit{f}_{\mathit{k}}})^{2} - \mathit{j}\frac{2\mathit{n}\varsigma\mathit{c}}{4\pi\mathit{f}_{\mathit{k}}}\right)}}.$ The constants $\mu_{0}$ and $\varepsilon_{0}$ represent the free-space permeability and permittivity respectively, $\mathit{n}$ denotes the index of refraction and $\varsigma$ represents the absorption coefficient of the reflecting medium. Note that, the traditional CSI estimation techniques such as LS/ MMSE cannot exploit the spatial-sparsity of the THz MIMO channel that occurs due to the higher link directionality. The virtual channel model for the THz system, that enables improved estimation, is described in the subsequent section.
\section{Sparse Channel Estimation Model for an SC-FDE THz System}
As discussed in Section II-A, the output vector at the RC for the $\mathit{m}$th block at the $\mathit{k}$th subcarrier is given by
\vspace{-5pt}
\begin{align}
\mathbf{{y}}_\mathit{m}[\mathit{k}] = \mathbf{\Phi}_{\mathit{m}}[\mathit{k}]\mathbf{h}[\mathit{k}] + \mathbf{{v}}_\mathit{m}[\mathit{k}]. \label{vec_out}
\end{align}
To derive an equivalent model \cite{rodriguez2018channel} for the $\mathit{k}$th subcarrier, we vertically stack the outputs $\mathbf{y}_\mathit{m}[\mathit{k}]$ for all the blocks, i.e,  $1\leq\mathit{m}\leq\textit{M}$
\vspace{-5pt}
\begin{align}
\underbrace{\begin{bmatrix}{\mathbf{y}}_{\text{1}}[k]\\ \vdots\\ {\mathbf{y}}_{\mathit{M}}[k]\end{bmatrix}}_{{\mathbf{y}_{\mathit{p}}}[k]} = \underbrace{ \begin{bmatrix} \mathbf{\Phi}_{\text{1}}[\mathit{k}]\\ \vdots\\ \mathbf{\Phi}_{\mathit{M}}[\mathit{k}]\end{bmatrix}}_{\mathbf{\Phi_{\mathit{p}}[\mathit{k}]}}\mathbf{h}[\mathit{k}] + \underbrace{\begin{bmatrix} \mathbf{v}_{\text{1}}[k]\\ \vdots\\ \mathbf{v}_{\mathit{M}}[k]\end{bmatrix}}_{\mathbf{v}_{\mathit{p}}[k]}, \label{con_frame}
\end{align}
where, $\mathbf{y}_{\mathit{p}}[\mathit{k}] \in \mathbb{C}^{\mathit{M}\mathit{N}_{\mathit{RF}}\times 1}$ represents the stacked pilot output vector, $\mathbf{\Phi}_{\mathit{p}}[\mathit{k}] \in \mathbb{C}^{\mathit{M}\mathit{N}_{\mathit{RF}}\times\mathit{N}_{\mathit{R}}\mathit{N}_{\mathit{T}}}$ denotes the stacked pilot sensing matrix and $\mathbf{v}_{\mathit{p}}[\mathit{k}]\in\mathbb{C}^{\mathit{M}\mathit{N}_{\mathit{RF}}\times1}$ represents the stacked noise corresponding to all the $\mathit{M}$ blocks for subcarrier $k$. Thus, the stacked system model at the combiner output is given by
\vspace{-5pt}
\begin{align}
\mathbf{y}_{\mathit{p}}[\mathit{k}] = \mathbf{\Phi}_{\mathit{p}}[\mathit{k}]\mathbf{h}[\mathit{k}] + \mathbf{v}_{\mathit{p}}[\mathit{k}]. \label{p_output}
\end{align}
At this point, one can formulate the traditional ZF and MMSE estimators for the wideband THz MIMO channel at each subcarrier as
\vspace{-5pt}
\begin{align}
\hat{\mathbf{h}}_{\textrm{LS}}[\mathit{k}] = (\mathbf{\Phi}_{\mathit{p}}^{\mathit{H}}[\mathit{k}]\mathbf{\Phi}_{\mathit{p}}[\mathit{k}])^{-1}\mathbf{\Phi}_{\mathit{p}}^{\mathit{H}}[\mathit{k}]\mathbf{y}_{\mathit{p}}[\mathit{k}] \label{ls_poutput}
\end{align}
\vspace{-20pt}
\begin{align}
\widehat{\mathbf{h}}_{\textrm{MMSE}}[\mathit{k}] = (\mathbf{R}_{\mathit{h}}^{-\textrm{1}}[\mathit{k}] + \mathbf{\Phi}_{\mathit{p}}^{\mathit{H}}[\mathit{k}]\mathbf{R}_{\mathit{v}_{\mathit{p}}}^{-\textrm{1}} \mathbf{\Phi}_{\mathit{p}}[\mathit{k}])^{-1}\mathbf{\Phi}_{\mathit{p}}^{\mathit{H}}[\mathit{k}]\mathbf{y}_{\mathit{p}}[\mathit{k}], \label{mmse_poutput}
\end{align}
where, the noise covariance matrix $\mathbf{R}_{\mathit{v}_{\mathit{p}}} = \sigma^2_v K\: \text{blkdiag}\left(\left\{\mathbf{W}^H_{RF,m}\mathbf{W}_{RF,m}\right\}_{m=1}^M\right)$ and $\mathbf{R}_{\mathit{v}_{\mathit{p}}} \in \mathbb{C}^{\mathit{M}\mathit{N}_{RF} \times \mathit{M}\mathit{N}_{RF}}$. Furthermore, $\mathbf{R}_{\mathit{h}}[\mathit{k}] = \mathbb{E}[\mathbf{h}[\mathit{k}]\mathbf{h}^{\mathit{H}}[\mathit{k}]] \in \mathbb{C}^{\mathit{N}_{R}\mathit{N}_{T}\times\mathit{N}_{R}\mathit{N}_{T}}$ denotes the channel covariance matrix. Note that the conventional LS and MMSE-based estimators require excessive pilot overheads in THz MIMO systems since they require an over-determined model, i.e, $\mathit{M}\mathit{N}_{\mathit{RF}} \geq \mathit{N}_{\mathit{T}}\mathit{N}_{\mathit{R}}$. This substantially degrades the spectral efficiency. Moreover, these orthodox techniques listed above do not leverage the specific properties of the THz channel. An important observation in this context is the fact that the directional nature of propagation in the THz band leads to an angularly sparse multipath channel \cite{srivastava2022hybrid}. In such scenarios, the recently developed sparse signal processing paradigm can yield excellent results for channel estimation in THz band hybrid MIMO systems, even in ill-posed scenarios, wherein we have $\mathit{M}\mathit{N}_{\mathit{RF}} \ll \mathit{N}_{\mathit{T}}\mathit{N}_{\mathit{R}}$. This feature may lead to substantial enhancement in the accuracy of channel estimation as well as bandwidth-efficiency. Therefore, the following section presents a sparse channel estimation model in THz band communication.
\vspace{-4pt}
\subsection{Sparse channel estimation model for a wideband THz hybrid MIMO system}
Let $\mathit{G}_{\mathit{T}}$ and $\mathit{G}_{\mathit{R}}$ represent the angular grid-sizes, which obey the relationship $(\mathit{G}_{\mathit{T}},\mathit{G}_{\mathit{R}}) \geq \textrm{max}(\mathit{N}_{\mathit{T}},\mathit{N}_{\mathit{R}})$. The transmit and receive angular grids $\Theta_T$ and $\Theta_R$, respectively, are created by considering the directional-cosines to be uniformly spaced between $-1$ to $1$, mathematically described as
\begin{align}
\Theta_{\mathit{T}} = \left\{\theta_{\mathit{t}}:\cos(\theta_{\mathit{t}}) = \frac{\textrm{2}}{\mathit{G}_{\mathit{T}}}(\mathit{t}-\textrm{1}) - 1, 1\leq\mathit{t}\leq\mathit{G}_{\mathit{T}} \right\},
\end{align}
\begin{align}
\Theta_{\mathit{R}} = \left\{\theta_{\mathit{r}}:\cos(\theta_{\mathit{r}}) = \frac{\textrm{2}}{\mathit{G}_{\mathit{R}}}(\mathit{r}-\textrm{1}) - 1, 1\leq\mathit{r}\leq\mathit{G}_{\mathit{R}} \right\}. \label{rec_ag}
\end{align}
Let $\mathbf{A}_{\mathit{T}}(\Theta_{\mathit{T}},\mathit{f}_k) \in \mathbb{C}^{\mathit{N}_{\mathit{T}}\times\mathit{G}_{\mathit{T}}}$ and $\mathbf{A}_{\mathit{R}}(\Theta_{\mathit{R}},\mathit{f}_k) \in \mathbb{C}^{\mathit{N}_{\mathit{R}}\times \mathit{G}_{\mathit{R}}}$ represent the dictionary matrices for the transmit and receive array manifold vectors using the angular grids $\Theta_{\mathit{T}}$ and $\Theta_{\mathit{R}}$ \cite{wang2018spatial}, which are modeled as
\begin{align}
\mathbf{A}_{\mathit{T}}(\Theta_{\mathit{T}},\mathit{f}_k) = [\tilde{\mathbf{a}}_{\mathit{T}}(\theta_{\textrm{1}},\mathit{f}_k),\tilde{\mathbf{a}}_{\mathit{T}}(\theta_{\textrm{2}},\mathit{f}_k),\cdots,\tilde{\mathbf{a}}_{\mathit{T}}(\theta_{\mathit{G}_{\mathit{T}}},\mathit{f}_k)], \label{trans_arrayres}
\end{align}
\begin{align}
\mathbf{A}_{\mathit{R}}(\Theta_{\mathit{R}},\mathit{f}_k) = [\tilde{\mathbf{a}}_{\mathit{R}}(\theta_{\textrm{1}},\mathit{f}_k),\tilde{\mathbf{a}}_{\mathit{R}}(\theta_{\textrm{2}},\mathit{f}_k),\cdots,\tilde{\mathbf{a}}_{\mathit{R}}(\theta_{\mathit{G}_{\mathit{R}}},\mathit{f}_k)]. \label{rec_arrayres}
\end{align}
Because of the particular selection of the AoD and AoA grids, the array response dictionary matrices \cite{srivastava2021data} satify the conditions 
\vspace{-8pt}
\begin{align}
\mathbf{A}_{\mathit{T}}(\Theta_{\mathit{T}},\mathit{f}_k)\mathbf{A}^{\mathit{H}}_{\mathit{T}}(\Theta_{\mathit{T}},\mathit{f}_k) = \frac{\mathit{G_{\mathit{T}}}}{\mathit{N}_{\mathit{T}}}\mathbf{I}_{\mathit{N}_{\mathit{T}}}, \notag
\end{align}
\vspace{-8pt}
\begin{align}
\mathbf{A}_{\mathit{R}}(\Theta_{\mathit{R}},\mathit{f}_k)\mathbf{A}^{\mathit{H}}_{\mathit{R}}(\Theta_{\mathit{R}},\mathit{f}_k)= \frac{\mathit{G_{\mathit{R}}}}{\mathit{N}_{\mathit{R}}}\mathbf{I}_{\mathit{N}_{\mathit{R}}}. \label{con_rec_arrayres}
\end{align}
It is noteworthy that the number of multipath components is substantially lower in a THz system due to the reflections, molecular absorption and free-space losses, as described earlier. This fact, coupled with the presence of only a few highly-directional beams, renders the THz MIMO channel $\mathbf{H}_{\mathit{b}}[k]$ angularly sparse in nature. Subsequently, the THz MIMO channel $\mathbf{H}[\mathit{k}]$ can be represented using the virtual channel model \cite{rodriguez2018channel} as 
\begin{align}
\mathbf{H}[\mathit{k}] = \mathbf{A}_{\mathit{R}}(\Theta_{\mathit{R}},\mathit{f}_{\mathit{k}}) {\mathbf{H}_{\mathit{b}}[\mathit{k}]}\mathbf{A}_{\mathit{T}}^{\mathit{H}}(\Theta_{\mathit{T}},\mathit{f}_{\mathit{k}}), \label{freq_beam_channel}
\end{align}
where $\mathbf{H}_{\mathit{b}}[\mathit{k}] \in \mathbb{C}^{\mathit{G}_{\mathit{R}} \times \mathit{G}_{\mathit{T}}}$ is the beamspace CFR matrix that pertains to $\mathbf{H}[\mathit{k}]$. Exploiting the property of the $\textrm{vec(.)}$ operator described in Section-I-C, the vectorized THz MIMO channel can be represented as
\begin{align}
\mathbf{h}[\mathit{k}] = \textrm{vec}(\mathbf{H}[\mathit{k}]) = \left [\mathbf{A}^{*}_{\mathit{T}}(\Theta_{\mathit{T}},\mathit{f}_{\mathit{k}}) \otimes \mathbf{A}_{\mathit{R}}(\Theta_{\mathit{R}},\mathit{f}_{\mathit{k}}) \right ]\mathbf{h}_{\mathit{b}}[\mathit{k}], \label{vec_channel_freq}
\end{align}
where $\mathbf{h}_{\mathit{b}}[\mathit{k}] = \textrm{vec}(\mathbf{H}_{\mathit{b}}[\mathit{k}]) \in \mathbb{C}^{\mathit{G}_{\mathit{R}}\mathit{G}_{\mathit{T}} \times 1}$. Let $\mathbf{\Psi}[\mathit{k}] = \left [\mathbf{A}^{*}_{\mathit{T}}(\Theta_{\mathit{T}},\mathit{f}_{\mathit{k}}) \otimes \mathbf{A}_{\mathit{R}}(\Theta_{\mathit{R}},\mathit{f}_{\mathit{k}}) \right ]$ denote the \textit{sparsifying-dictionary}, where $\mathbf{\Psi}[\mathit{k}] \in \mathbb{C}^{\mathit{N}_{\mathit{R}} \mathit{N}_{\mathit{T}}  \times \mathit{G}_{\mathit{R}} \mathit{G}_{\mathit{T}}}$. Thus, upon substituting \eqref{vec_channel_freq} in to \eqref{p_output}, the expression for the output vector can be recast as
\begin{align}
\mathbf{y}_{\mathit{p}}[\mathit{k}] = \underbrace{\mathbf{\Phi}_{\mathit{p}}[\mathit{k}]\mathbf{\Psi}[\mathit{k}]}_{\tilde{\mathbf{\Psi}}_{\mathit{p}}[\mathit{k}]}\mathbf{h}_{\mathit{b}}[k] + \mathbf{v}_{\mathit{p}}[\mathit{k}], \label{vec_pilot_beam}
\end{align}
where $\tilde{\mathbf{\Psi}}_{\mathit{p}}[\mathit{k}] \in \mathbb{C}^{\mathit{M}\mathit{N}_{\mathit{RF}} \times \mathit{G}_{\mathit{R}}\mathit{G}_{\mathit{T}}}$ represents an equivalent sensing matrix. It is noteworthy that other sparse estimation techniques, such as Basis-Pursuit \cite{ekanadham2011recovery}, FOCUSS \cite{cotter2005sparse}, and OMP \cite{kulkarni2017low}, suffer from structural and convergence-related issues. For instance, in the case of Basis-Pursuit, there is no guarantee that the sparsest solution aligns with the global minimum of its cost function, leading to structural errors. Similarly, the FOCUSS algorithm frequently converges to suboptimal local minima, resulting in convergence errors. Additionally, the OMP algorithm presents challenges; a low cessation criterion leads to a larger number of iterations and more non-zero components in the sparse vector, causing structural errors. Conversely, a high cessation criterion results in fewer iterations, failing to capture all the dominant modes of the channel and causing convergence errors. By contrast, the proposed Bayesian learning methods assure global convergence with a high probability due to the inherent attributes of their cost function and the use of the EM algorithm. Furthermore, in a noiseless scenario, the global minimum of the Bayesian learning cost function corresponds to the sparsest solution \cite{wipf2004sparse}. Even in a noisy scenario, all the local minima are sparse and have improved representation over non-sparse solutions. The next section describes the framework that utilizes Bayesian learning to estimate $\mathbf{h}_b[k]$, that intends to exploit the sparsity of the THz MIMO channel as described above.
\section{PA-BL for Sparse Channel Estimation in THz MIMO Systems}
This section introduces a Pilot Aided-Bayesian Learning (PA-BL) technique for CSI estimation in dual-wideband channel. This can also be applied for the case of an under-determined system, i.e, $\mathit{M}\mathit{N}_{\mathit{RF}}\ll\mathit{G}_{\mathit{T}}\mathit{G}_{\mathit{R}}$ \cite{srivastava2021data}. At first, the parameterized Gaussian prior $\mathit{f}(\mathbf{h}_{\mathit{b}}[\mathit{k}];\mathbf{\Gamma}_{\mathit{k}})$ is allocated to the $\mathit{k}$th pilot subcarrier for the sparse vector $\mathbf{h}_{\mathit{b}}[\mathit{k}]$ \cite{wipf2004sparse}, which is given by
\vspace{-5pt}
 \begin{align}
\mathit{f}(\mathbf{h}_{\mathit{b}}[\mathit{k}];\mathbf{\Gamma}_{\mathit{k}}) = \prod_{\mathit{i}=1}^{\mathit{G}_{\mathit{R}}\mathit{G}_{\mathit{T}}}(\pi\gamma_{\mathit{k},\mathit{i}})^{-1}\textrm{exp}\left (-\frac{|\mathbf{h}_{\mathit{b}}[\mathit{k}](\mathit{i})|^{2}}{\gamma_{\mathit{k},\mathit{i}}}\right), \label{pra_prior}
\end{align}
where $\gamma_{\mathit{k},\mathit{i}}$ represents $i$th hyper-parameter for the $\mathit{k}$th subcarrier; $1\leq\mathit{i}\leq\mathit{G}_{\mathit{R}}\mathit{G}_{\mathit{T}}$, and the matrix $\mathbf{\Gamma}_{\mathit{k}}\in\mathbb{R}^{\mathit{G}_{\mathit{R}}\mathit{G}_{\mathit{T}}\times\mathit{G}_{\mathit{R}}\mathit{G}_{\mathit{T}}}$ is the diagonal matrix of hyper-parameters is given by $\mathbf{\Gamma}_{\mathit{k}} = \textrm{diag} (\gamma_{\mathit{k},\textrm{1}},\gamma_{\mathit{k},\textrm{2}},\cdots, \gamma_{\mathit{k},\mathit{G}_{\mathit{R}}\mathit{G}_{\mathit{T}}}).$ The MMSE estimate of $\hat{\mathbf{h}}_{\mathit{b}}[\mathit{k}]$ is given by
 \begin{align}
\hat{\mathbf{h}}_{\mathit{b}}[\mathit{k}] = \left(\tilde{\mathbf{\Psi}}_{\mathit{p}}^{\mathit{H}}[\mathit{k}] \mathbf{R}_{\mathit{v}_{\mathit{p}}}^{-1} \tilde{\mathbf{\Psi}}_{\mathit{p}}[\mathit{k}] + \mathbf{\Gamma}_{\mathit{k}}^{-1} \right)^{-1} \tilde{\mathbf{\Psi}}_{\mathit{p}}^{\mathit{H}}[\mathit{k}] \mathbf{R}_{\mathit{v}_{\mathit{p}}}^{-1}\mathbf{y}_{\mathit{p}}[\mathit{k}]. \label{mmse_beam}
\end{align}
It can be noted that MMSE relies on the hyper-parameter matrix $\mathbf{\Gamma}_{\mathit{k}}$. Therefore, estimating $\mathbf{\Gamma}_{\mathit{k}}$ holds the key toward determining the sparse channel $\hat{\mathbf{h}}_{\mathit{b}}[\mathit{k}]$. The log-likelihood $\textrm{log}\left[\mathit{f}\left(\mathbf{y}_{\mathit{p}}[\mathit{k}];\mathbf{\Gamma}_{\mathit{k}}\right)\right]$ of the hyper-parameter matrix $\mathbf{\Gamma}_{\mathit{k}}$ can be formulated as
\begin{align}
\textrm{log}\left[\mathit{f}\left(\mathbf{y}_{\mathit{p}}[\mathit{k}];\mathbf{\Gamma}_{\mathit{k}}\right)\right] = \mathit{c}_{\textrm{1}} - & \textrm{log}\left[\textrm{det}\left(\mathbf{R}_{\mathbf{y}_{\mathit{p}}}[\mathit{k}]\right)\right] - \notag \\ & \mathbf{y}_{\mathit{p}}^{\mathit{H}}[\mathit{k}]\mathbf{R}_{\mathbf{y}_{\mathit{p}}}^{\textrm{-1}}[\mathit{k}]
\mathbf{y}_{\mathit{p}}[\mathit{k}], \label{PA-BL_evi}
\end{align}
where $\mathit{c}_{\textrm{1}} = -\mathit{M}\mathit{N}_{\mathit{RF}}\textrm{log}(\pi)$ and the pilot covariance matrix $\mathbf{R}_{\mathbf{y}_{\mathit{p}}}[\mathit{k}]\in\mathbb{C}^{\mathit{M}\mathit{N}_{\mathit{RF}}\times\mathit{M}\mathit{N}_{\mathit{RF}}}$ is given by
\begin{align}
\mathbf{R}_{\mathbf{y}_{\mathit{p}}}[\mathit{k}] = \mathbb{E}\left\{\mathbf{y}_{\mathit{p}}[\mathit{k}]\mathbf{y}_{\mathit{p}}^{\mathit{H}}[\mathit{k}]\right\}. \label{pilot_cov}
\end{align}
Substituting equation \eqref{vec_pilot_beam} in to \eqref{pilot_cov} yields the pilot covariance matrix in the form of
\begin{align}
\mathbf{R}_{\mathbf{y}_{\mathit{p}}}[\mathit{k}] = \mathbf{R}_{\mathit{v}_{\mathit{p}}} + \tilde{\boldsymbol{\Psi}}_{\mathit{p}}[\mathit{k}]\mathbf{\Gamma}_{\mathit{k}}\tilde{\boldsymbol{\Psi}}_{\mathit{p}}^{\mathit{H}}[\mathit{k}], \label{noise_pilot_cov}
\end{align}
where $\mathbf{v}_{\mathit{p}}[\mathit{k}]\sim\mathcal{CN}(\mathbf{0},\mathbf{R}_{\mathit{v}_{\mathit{p}}})$. Maximization of the log-likelihood with respect to $\mathbf{\Gamma}_{\mathit{k}}$ is mathematically intractable. Thus, the Expectation Maximization (EM) algorithm is a suitable choice for iterative maximization, as it ensures convergence to a local optimum \cite{mishra2017sparse}. The EM-based procedure of determining the sparse channel estimate is described below. Let the estimate of the $\mathit{i}$th hyperparameter at the $\mathit{k}$th subcarrier obtained from the $(\mathit{j} - 1)$ st EM iteration be represented as $\hat{\gamma}_{\mathit{k},\mathit{i}}^{\left(\mathit{j}-1\right)}$ and $\hat{\mathbf{\Gamma}}_{\mathit{k}}^{(\mathit{j}-1)}$ denotes the hyperparameter matrix at the $\mathit{k}$th subcarrier, which is further defined as $\hat{\mathbf{\Gamma}}_{\mathit{k}}^{(\mathit{j}-1)} = \textrm{diag}\left(\hat{\gamma}_{\mathit{k},1}^{(\mathit{j}-1)}, \hat{\gamma}_{\mathit{k},2}^{(\mathit{j}-1)},\cdots,\hat{\gamma}_{\mathit{k},\mathit{G}_{\mathit{R}}\mathit{G}_{\mathit{T}}}^{(\mathit{j}-1)}\right).$ The steps involved in updating the estimate $\hat{\mathbf{\Gamma}}_{\mathit{k}}^{\left(\mathit{j}\right)}$ in the $\mathit{j}$th iteration of the EM algorithm can be outlined as follows. Let the $\mathit{i}$th hyper-parameter estimate obtained in iteration $(j-1)$st for the $\mathit{k}$th subcarrier be denoted by $\hat{\gamma}_{\mathit{k},\mathit{i}}^{(\mathit{j}-1)}$. The hyper-parameter update $\hat{\gamma}_{\mathit{k},\mathit{i}}^{(\mathit{j})}$ in the $\mathit{j}$th EM-iteration is obtained by maximizing the log-likelihood function $\mathcal{L}\left(\mathbf{\Gamma}_{\mathit{k}}|\hat{\mathbf{\Gamma}}_{\mathit{k}}^{(\mathit{j}-1)}\right)$, defined as
\begin{align}
\mathcal{L}\left (\mathbf{\Gamma}_{\mathit{k}}|\hat{\mathbf{\Gamma}}_{\mathit{k}}^{(\mathit{j}-1)}\right) = \mathbb{E}_{\mathbf{h}_{\mathit{b}}[\mathit{k}]\mathbf{|}\mathbf{y}_{\mathit{p}}[\mathit{k}];\hat{\mathbf{\Gamma}}_{\mathit{k}}^{(\mathit{j}-1)}} \left\{{\mathrm{log} \ \mathit{f}\left(\mathbf{y}_{\mathit{p}}[\mathit{k}],\mathbf{h}_{\mathit{b}}[\mathit{k}];\mathbf{\Gamma_{\mathit{k}}}\right)} \right\}. \label{log_hyper}
\end{align}
The conditional expectation of the log-likelihood function $\mathcal{L}\left(\mathbf{\Gamma}_\mathit{k}|\hat{\mathbf{\Gamma}}_{\mathit{k}}^{\left(\mathit{j}- 1\right)}\right)$ in the E-step is given 
\begin{align}
\mathcal{L}\left(\mathbf{\Gamma}_\mathit{k}|\hat{\mathbf{\Gamma}}_{\mathit{k}}^{\left(\mathit{j}- 1\right)}\right) = & \mathbb{E}_{\mathbf{h}_{\mathit{b}}[\mathit{k}]\mathbf{|}\mathbf{y}_{\mathit{p}}[\mathit{k}];\hat{\mathbf{\Gamma}}_{\mathit{k}}^{(\mathit{j}-1)}} \left\{{\textrm{log} \ \mathit{f}\left(\mathbf{y}_{\mathit{p}}[\mathit{k}],\mathbf{h}_{\mathit{b}}[\mathit{k}];\mathbf{\Gamma_{\mathit{k}}}\right)} \right\} \notag \\
= & \mathbb{E}_{\mathbf{h}_{\mathit{b}}[\mathit{k}]\mathbf{|}\mathbf{y}_{\mathit{p}}[\mathit{k}];\hat{\mathbf{\Gamma}}_{\mathit{k}}^{(\mathit{j}-1)}}\left\{\textrm{log}\left[\mathit{f}(\mathbf{y}_{\mathit{p}}[\mathit{k}]|\mathbf{h}_{\mathit{b}}[\mathit{k}])\right]\right\} + \notag \\ & \mathbb{E}_{\mathbf{h}_{\mathit{b}}[\mathit{k}]\mathbf{|}\mathbf{y}_{\mathit{p}}[\mathit{k}];\hat{\mathbf{\Gamma}}_{\mathit{k}}^{(\mathit{j}-1)}}\left\{\textrm{log}\left[\mathit{f}(\mathbf{h}_{\mathit{b}}[\mathit{k}];\mathbf{\Gamma}_{\mathit{k}})\right]\right\}. \label{pilot_log_exp}
\end{align}
The expression enclosed within the first expectation operator in the aforementioned equation can be simplified as 
\begin{equation}
\begin{aligned}
&\textrm{log} \ \mathit{f}(\mathbf{y}_{\mathit{p}}[\mathit{k}]|\mathbf{h}_{\mathit{b}}[\mathit{k}]) = -\mathit{M}\mathit{N}_{\textrm{RF}} \ \textrm{log}(\pi) - \textrm{log}\left[\textrm{det}(\mathbf{R}_{\mathit{v}_{\mathit{p}}})\right] - \cr &\left(\mathbf{y}_{\mathit{p}}[\mathit{k}] - \tilde{\boldsymbol{\Psi}}_{\mathit{p}}[\mathit{k}]\mathbf{h}_{\mathit{b}}[\mathit{k}]\right)^{\mathit{H}}\mathbf{R}_{\mathit{v}_{\mathit{p}}}^{-1}\left(\mathbf{y}_{\mathit{p}}[\mathit{k}] - \tilde{\boldsymbol{\Psi}}_{\mathit{p}}[\mathit{k}]\mathbf{h}_{\mathit{b}}[\mathit{k}]\right). \label{exp_yb_hb}
\end{aligned}
\end{equation}
It can be seen that the above expression is independent of the hyper-parameter matrix  $\mathbf{\Gamma}_{\mathit{k}}$, which implies that it can be overlooked in the ensuing M-step of the CSI estimation algorithm. Thus, the equivalent optimization problem of determining the beamspace CSI estimate can be formulated
\begin{equation}
\begin{aligned}
\hat{\boldsymbol{\Gamma}}_{\mathit{k}}^{\left(\mathit{j}\right)} = \underset{\mathbf{\Gamma}_{\mathit{k}}}{\textrm{arg max}} \ \mathbb{E}_{\mathbf{h}_{\mathit{b}}[\mathit{k}]|\mathbf{y}_{\mathit{p}}[\mathit{k}];\hat{\mathbf{\Gamma}}_{\mathit{k}}^{\left(\mathit{j}-1\right)}}\left\{\textrm{log}\left[\mathit{f}\left(\mathbf{h}_{\mathit{b}}[\mathit{k}];\boldsymbol{\Gamma}_{\mathit{k}}\right)\right]\right\}. \label{sub_hyper_esti}
\end{aligned}
\end{equation}
On substituting $\mathit{f}(\mathbf{h}_{\mathit{b}}[\mathit{k}];\mathbf{\Gamma}_{\mathit{k}})$ from equation \eqref{pra_prior} into the optimization objective, one can observe that the resultant maximization problem can be separated in reference to the hyper-parameters, and the individual optimization problems are given as
\vspace{-8pt}
\begin{align}
\hat{\gamma}_{\mathit{k},\mathit{i}}^{\left(\mathit{j}\right)} = \underset{\gamma_{\mathit{k},\mathit{i}}}{\textrm{arg max}}\left[-\textrm{log}(\gamma_{\mathit{k},\mathit{i}}) - \frac{\mathbb{E}_{\mathbf{h}_{\mathit{b}}[\mathit{k}]|\mathbf{y}_{\mathit{p}}[\mathit{k}];\hat{\boldsymbol{\Gamma}}_{\mathit{k}}^{(\mathit{j}-1)}}\left\{|\mathbf{h}_{\mathit{b}}[\mathit{k}](\mathit{i})|^{2}\right\}}{\gamma_{\mathit{k},\mathit{i}}}\right]. \label{opti_hyper}
\end{align}
On solving the aforementioned equation, the estimates $\hat{\gamma}_{\mathit{k},\mathit{i}}^{\left(\mathit{j}\right)}$ are determined as
\begin{align}
\hat{\gamma}_{\mathit{k},\mathit{i}}^{\left(\mathit{j}\right)} = \mathbb{E}_{\mathbf{h}_{\mathit{b}}[\mathit{k}]|\mathbf{y}_{\mathit{p}}[\mathit{k}];\hat{\boldsymbol{\Gamma}}_{\mathit{k}}^{(\mathit{j}-1)}}\left\{|\mathbf{h}_{\mathit{b}}[\mathit{k}](\mathit{i})|^{2}\right\}, \label{soln_opti_hyper}
\end{align}
where, the \textit{a posteriori} distribution of $\mathbf{h}_{\mathit{b}}[\mathit{k}]$, required to simplify the conditional expectation $\mathbb{E}_{\mathbf{h}_b[k]|\mathbf{y}_p[k];\hat{\mathbf{\Gamma}}^{(j-1)}_k}\left\{.\right\}$ above, can be determined as
\vspace{-3pt}
\begin{align}
f\left(\mathbf{h}_{\mathit{b}}[\mathit{k}]|\mathbf{y}_{\mathit{p}}[\mathit{k}];\hat{\boldsymbol{\Gamma}}_{\mathit{k}}^{\left(\mathit{j}-1\right)}\right)= \mathcal{CN}\left(\boldsymbol{\mu}_{\mathit{b}}^{(\mathit{j})}[\mathit{k}],\mathbf{R}_{\mathit{b}}^{(\mathit{j})}[\mathit{k}]\right). \label{pd_mean_cov}
\end{align}
The quantity $\boldsymbol{\mu}_{\mathit{b}}^{(\mathit{j})}[\mathit{k}]\in\mathbb{C}^{\mathit{G}_{\mathit{R}}\mathit{G}_{\mathit{T}}\times1}$ represents the \textit{a posteriori} mean, while the quantity $\mathbf{R}_{\mathit{b}}^{(\mathit{j})}[\mathit{k}]\in\mathbb{C}^{\mathit{G}_{\mathit{R}}\mathit{G}_{\mathit{T}}\times\mathit{G}_{\mathit{R}}\mathit{G}_{\mathit{T}}}$ represents the covariance matrix. These expressions are given below
\vspace{-3pt}
\begin{equation}
\boldsymbol{\mu}_{\mathit{b}}^{\left(\mathit{j}\right)}[\mathit{k}] = \mathbf{R}_{\mathit{b}}^{\left(\mathit{j}\right)}[\mathit{k}]\tilde{\mathbf{\Psi}}_{\mathit{p}}^{\mathit{H}}[\mathit{k}]\mathbf{R}_{\mathit{v}_{\mathit{p}}}^{-\textrm{1}}\mathbf{y}_{\mathit{p}}[\mathit{k}], \notag
\end{equation}
\begin{equation}
\mathbf{R}_{\mathit{b}}^{\left(\mathit{j}\right)}[\mathit{k}] = \left[\tilde{\mathbf{\Psi}}_{\mathit{p}}^{\mathit{H}}[\mathit{k}]\mathbf{R}_{\mathit{v}_{\mathit{p}}}^{-\textrm{1}}\tilde{\mathbf{\Psi}}_{\mathit{p}}[\mathit{k}] + \left(\hat{\mathbf{\Gamma}}_{\mathit{k}}^{\left(\mathit{j} - 1\right)}\right)^{-\textrm{1}}\right]^{-\textrm{1}}. \label{cov_esti}
\end{equation}
Using the \textit{a posteriori} PDF from \eqref{pd_mean_cov}, the expression in \eqref{soln_opti_hyper} for the hyper-parameter estimate reduces to
\begin{align}
\hat{\gamma}_{\mathit{k},\mathit{i}}^{\left(\mathit{j}\right)} = \mathbf{R}_{\mathit{b}}^{\left(\mathit{j}\right)}[\mathit{k}](\mathit{i},\mathit{i}) + |\boldsymbol{\mu }_{\mathit{b}}^{\left(\mathit{j}\right)}[\mathit{k}](\mathit{i})|^{2}. \label{em_esti}
\end{align}
\begin{algorithm}[t!]
\caption{PA-BL for sparse channel estimation in THz MIMO systems}\label{algo_SBL}
\textbf{Input:} Pilot output $\mathbf{y}_{p}[\mathit{k}]$, equivalent sensing matrix $\tilde{\mathbf{\Psi}}_{p}[\mathit{k}]$, noise covariance $\mathbf{R}_{\mathit{v_{p}}}$, dictionary matrices $\mathbf{A}_{\mathit{R}}(\Theta_{\mathit{R}},\mathit{f}_{\mathit{k}})$ and $\mathbf{A}_{\mathit{T}}(\Theta_{\mathit{T}},\mathit{f}_{\mathit{k}})$, stopping parameters $\epsilon$ and $\mathit{K}_{\textrm{max}}$\\
\textbf{Initialization:} $\gamma_{\mathit{k},\mathit{i}}^{\left(0\right)} = \textrm{1}, \forall \ \textrm{1}\leq\mathit{i}\leq\mathit{G}_{\mathit{R}}\mathit{G}_{\mathit{T}}$ where $\textrm{1}\leq\mathit{k}\leq\mathit{K}$ $\Rightarrow \hat{\mathbf{\Gamma}}_{\mathit{k}}^{\left(\textrm{0}\right)} = \mathbf{I}_{\mathit{G}_{\mathit{R}}\mathit{G}_{\mathit{T}}}$, $\hat{\mathbf{\Gamma}}_{\mathit{k}}^{\left(-1\right)} = \mathbf{0}$ and counter $\mathit{j} = \textrm{0}$ \\
\textbf{while} $\left(\parallel \hat{\mathbf{\Gamma}}_{\mathit{k}}^{\left(\mathit{j}\right)} - \hat{\mathbf{\Gamma}}_{\mathit{k}}^{\left(\mathit{j}-1\right)}\parallel_{\mathit{F}} \ > \epsilon \ \textrm{and} \  \mathit{j} < \mathit{K}_{\textrm{max}}\right)$ \textbf{do}
\begin{enumerate}
\item $j \leftarrow j+1$
\item \textbf{E-step: } Calculate the \textit{a posteriori} mean and covariance
\begin{equation}
\mathbf{R}_{\mathit{b}}^{\left(\mathit{j}\right)}[\mathit{k}] = \left[\tilde{\mathbf{\Psi}}_{\mathit{p}}^{\mathit{H}}[\mathit{k}]\mathbf{R}_{\mathit{v}_{\mathit{p}}}^{-\textrm{1}}\tilde{\mathbf{\Psi}}_{\mathit{p}}[\mathit{k}] + \left(\hat{\mathbf{\Gamma}}_{\mathit{k}}^{\left(\mathit{j} - 1\right)}\right)^{-\textrm{1}}\right]^{-\textrm{1}}\notag
\end{equation}
\begin{equation}
\boldsymbol{\mu}_{\mathit{b}}^{\left(\mathit{j}\right)}[\mathit{k}] = \mathbf{R}_{\mathit{b}}^{\left(\mathit{j}\right)}[\mathit{k}]\tilde{\mathbf{\Psi}}_{p}^{\mathit{H}}[\mathit{k}]\mathbf{R}_{\mathit{v_{p}}}^{-\textrm{1}}\mathbf{y}_{p}[\mathit{k}]\notag
\end{equation}
\item \textbf{M-step:} Update the hyperparameters\\ 
\phantom{beta}  \textbf{for} $\mathit{i} = 1,\cdots, \mathit{G}_{\mathit{R}}\mathit{G}_{\mathit{T}}$ \textbf{do}
\begin{equation}
\hat{\gamma}_{\mathit{k},\mathit{i}}^{\left(\mathit{j}\right)} = \mathbf{R}_{\mathit{b}}^{\left(\mathit{j}\right)}[\mathit{k}](\mathit{i},\mathit{i}) + |\boldsymbol{\mu}_{\mathit{b}}^{\left(\mathit{j}\right)}[\mathit{k}](\mathit{i})|^{2}\notag
\end{equation}
\phantom{beta} \textbf{end for}
\end{enumerate}
\textbf{end while}\\
$\hat{\mathbf{h}}_{\mathit{b}}[\mathit{k}] = \boldsymbol{\mu}_{\mathit{b}}^{\left(\mathit{j}\right)}[\mathit{k}]$ \\
\textbf{Output:} $\widehat{\mathbf{H}}_{\textrm{PA-BL}}[\mathit{k}] = \mathbf{A}_{\mathit{R}}(\Theta_{\mathit{R}},\mathit{f}_{\mathit{k}})\textrm{vec}^{-1}(\widehat{\mathbf{h}}_{\mathit{b}}[\mathit{k}])\mathbf{A}_{\mathit{T}}^{\mathit{H}}(\Theta_{\mathit{T}},\mathit{f}_{\mathit{k}})$
\end{algorithm}
Algorithm 1 provides an overview of the sparse channel estimation technique for THz hybrid MIMO systems, employing the PA-BL framework \cite{srivastava2022hybrid}. The procedure is repeated until the hyperparameter estimate converges, i.e, $\parallel\hat{\mathbf{\Gamma}}_{\mathit{k}}^{\left(\mathit{j}\right)}-\hat{\mathbf{\Gamma}}_{\mathit{k}}^{\left(\mathit{j}-1\right)}\parallel_{\mathit{F}}^{\textrm{2}} \ < \ \epsilon_{\textrm{0}}$ or the number of iterations reached its maximum affordable value $\mathit{K}_{\textrm{max}}$. The PA-BL based channel estimate after convergence is given by
\begin{align}
\hat{\mathbf{h}}_{\mathit{b}}[\mathit{k}] = \boldsymbol{\mu}_{\mathit{b}}^{\left(\mathit{j}\right)}[\mathit{k}]. \label{beam_channel_mean}
\end{align}
\subsection{Bayesian Cram{\'e}r-Rao lower bound for pilot-aided learning}
Consider the parameterized Gaussian prior assigned to the $\mathit{i}$th hyper-parameter of the $\mathit{k}$th pilot subcarrier corresponding to $\mathbf{h}_{\mathit{b}}[\mathit{k}]$, as mentioned in \eqref{pra_prior}. The corresponding PDF is given by
\vspace{-4pt}
\begin{align}
\mathit{f}(\mathbf{h}_{\mathit{b}}[\mathit{k}];\mathbf{\Gamma}_{\mathit{k}}) = \frac{1}{(\pi)^{\mathit{G}_{\mathit{R}}\mathit{G}_{\mathit{T}}} \ \textrm{det}(\mathbf{\Gamma}_{\mathit{k}})}\textrm{exp}\left(-\mathbf{h}_{\mathit{b}}^{\mathit{H}}[\mathit{k}]\mathbf{\Gamma}_{\mathit{k}}^{-1}\mathbf{h}_{\mathit{b}}[\mathit{k}]\right). \label{pra_prior_mat}
\end{align}
The log-likelihood of the beamspace channel $\mathbf{h}_{\mathit{b}}[\mathit{k}]$ is obtained via applying the log function to both sides of the above equation, yielding
\begin{align}
\mathcal{L}\left(\mathit{f}\left[\mathbf{h}_{\mathit{b}}[\mathit{k}];\mathbf{\Gamma}_{\mathit{k}}\right]\right) = & -\mathit{G}_{\mathit{R}}\mathit{G}_{\mathit{T}}\textrm{log}(\pi)-\textrm{log}\left[\textrm{det}(\mathbf{\Gamma}_{\mathit{k}})\right]- \notag \\ & \mathbf{h}_{\mathit{b}}^{\mathit{H}}[\mathit{k}]\mathbf{\Gamma}_{\mathit{k}}^{-1}\mathbf{h}_{\mathit{b}}[\mathit{k}]. \label{sol_pra_prior_mat}
\end{align}
The conditional PDF of the pilot output vector $\mathbf{y}_{\mathit{p}}[\mathit{k}]$, as defined in equation \eqref{vec_pilot_beam}, is given by the complex normal distribution $\mathcal{CN}\left(\tilde{\mathbf{\Psi}}_{\mathit{p}}[\mathit{k}]\mathbf{h}_{\mathit{b}}[\mathit{k}], \mathbf{R}_{\mathit{v}_{\mathit{p}}}\right)$. Hence, the expression of the conditional PDF is
\begin{align}
\mathit{f}\left(\mathbf{y}_{\mathit{p}}[\mathit{k}]|\mathbf{h}_{\mathit{b}}[\mathit{k}]\right) & = \vartheta[\mathit{k}]\textrm{exp}\big(-\left(\mathbf{y}_{\mathit{p}}[\mathit{k}]-\tilde{\boldsymbol{\Psi}}_{\mathit{p}}[\mathit{k}]\mathbf{h}_{\mathit{b}}[\mathit{k}]\right)^{\mathit{H}} \notag \\ & \mathbf{R}_{\mathit{v}_{\mathit{p}}}^{-1}\big(\mathbf{y}_{\mathit{p}}[\mathit{k}]-\tilde{\boldsymbol{\Psi}}_{\mathit{p}}[\mathit{k}]\mathbf{h}_{\mathit{b}}[\mathit{k}]\big)\big), \label{lik_pilot}
\end{align}
where $\vartheta[\mathit{k}] = \frac{1}{\left(\pi\right)^{\mathit{M}\mathit{N}_{\mathit{RF}}}|\mathbf{R}_{\mathit{v}_{\mathit{p}}}|}$. Taking the logarithm of both sides of the equation above gives us
\begin{equation}
\begin{aligned}
&\mathcal{L}\left(\mathbf{y}_{\mathit{p}}[\mathit{k}]|\mathbf{h}_{\mathit{b}}[\mathit{k}]\right) = -\mathit{M}\mathit{N}_{RF} \ \textrm{log}(\pi) - \textrm{log}\left[\textrm{det}(\mathbf{R}_{\mathit{v}_{\mathit{p}}})\right]- \cr &\left(\mathbf{y}_{\mathit{p}}[\mathit{k}] - \tilde{\mathbf{\Psi}}_{\mathit{p}}[\mathit{k}]\mathbf{h}_{\mathit{b}}[\mathit{k}]\right)^{\mathit{H}}\mathbf{R}_{\mathit{v}_{\mathit{p}}}^{-1}\left(\mathbf{y}_{\mathit{p}}[\mathit{k}] - \tilde{\mathbf{\Psi}}_{\mathit{p}}[\mathit{k}]\mathbf{h}_{\mathit{b}}[\mathit{k}]\right). \label{log_lik_pilot}
\end{aligned}
\end{equation}
Let $\mathbf{J}[\mathit{k}]\in\mathbb{C}^{\mathit{G}_{\mathit{R}}\mathit{G}_{\mathit{T}}\times\mathit{G}_{\mathit{R}}\mathit{G}_{\mathit{T}}}$ denote the Bayesian Fisher Information Matrix (FIM), formulated as
\begin{align}
\mathbf{J}[\mathit{k}] = \mathbf{J}_{\mathit{p}}[\mathit{k}] + \mathbf{J}_{\mathit{b}}[\mathit{k}], \label{FIM}
\end{align}
where $\mathbf{J}_{\mathit{p}}[\mathit{k}]\in\mathbb{C}^{\mathit{G}_{\mathit{R}}\mathit{G}_{\mathit{T}}\times\mathit{G}_{\mathit{R}}\mathit{G}_{\mathit{T}}}$ denotes the FIM associated with the pilot output $\mathbf{y}_p[k]$, while $\mathbf{J}_{\mathit{b}}[\mathit{k}]\in\mathbb{C}^{\mathit{G}_{\mathit{R}}\mathit{G}_{\mathit{T}}\times\mathit{G}_{\mathit{R}}\mathit{G}_{\mathit{T}}}$ represents the FIM corresponding to the \textit{a prior} information of the beamspace CSI $\mathbf{h}_b[k]$. The above quantities can be formulated as
\begin{align}
\mathbf{J}_{\mathit{p}}[\mathit{k}] = -\mathbb{E}_{\mathbf{y}_{\mathit{p}}[\mathit{k}],\mathbf{h}_{\mathit{b}}[\mathit{k}]}\left\{\frac{\partial^{2}\mathcal{L}(\mathbf{y}_{\mathit{p}}[\mathit{k}]|\mathbf{h}_{\mathit{b}}[\mathit{k}])}{\partial\mathbf{h}_{\mathit{b}}[\mathit{k}]\partial\mathbf{h}_{\mathit{b}}^{\mathit{H}}[\mathit{k}]}\right\} \notag,
\end{align}
\begin{align}
\mathbf{J}_{\mathit{b}}[\mathit{k}] = -\mathbb{E}_{\mathbf{h}_{\mathit{b}}[\mathit{k}]}\left\{\frac{\partial^{2}\mathcal{L}(\mathbf{h}_{\mathit{b}}[\mathit{k}];\mathbf{\Gamma}_{\mathit{k}})}{\partial\mathbf{h}_{\mathit{b}}[\mathit{k}]\partial\mathbf{h}_{\mathit{b}}^{\mathit{H}}[\mathit{k}]}\right\}. \label{beam_FIM}
\end{align}
Upon using the conditional PDF in \eqref{log_lik_pilot}, the FIM $\mathbf{J}_{\mathit{p}}[\mathit{k}]$ can be derived as
\begin{align}
\mathbf{J}_{\mathit{p}}[\mathit{k}] = \tilde{\mathbf{\Psi}}_{\mathit{p}}^{\mathit{H}}[\mathit{k}]\mathbf{R}_{\mathit{v}_{\mathit{p}}}^{-1}\tilde{\mathbf{\Psi}}_{\mathit{p}}[\mathit{k}]. \label{sol_pilot_FIM}
\end{align}
Similarly, the FIM associated with $\mathbf{h}_{\mathit{b}}[\mathit{k}]$ is obtained by using the \textit{a prior} PDF from equation \eqref{pra_prior_mat} in equation \eqref{beam_FIM}. The corresponding expression for $\mathbf{J}_{\mathit{b}}[\mathit{k}]$ can be formulated as 
\begin{align}
\mathbf{J}_{\mathit{b}}[\mathit{k}] = \hat{\mathbf{\Gamma}}_{\mathit{k}}^{-1}. \label{sol_beam_FIM}
\end{align}
Using the expressions in \eqref{sol_pilot_FIM} and \eqref{sol_beam_FIM}, the Bayesian FIM is given by
\begin{align}
\mathbf{J}[\mathit{k}] = \tilde{\mathbf{\Psi}}_{\mathit{p}}^{\mathit{H}}[\mathit{k}]\mathbf{R}_{\mathit{v}_{\mathit{p}}}^{-1}\tilde{\mathbf{\Psi}}_{\mathit{p}}[\mathit{k}] + \hat{\mathbf{\Gamma}}_{\mathit{k}}^{-1}. \label{sol_FIM}
\end{align}
Therefore, the BCRLB that pertains to the MSE of the estimate $\hat{\mathbf{h}}_{\mathit{b}}[\mathit{k}]$ can be expressed as
\begin{align}
\begin{split}
\textrm{MSE}(\hat{\mathbf{h}}_{\mathit{b}}[\mathit{k}]) & = \mathbb{E}\left\{\parallel\hat{\mathbf{h}}_{\mathit{b}}[\mathit{k}]-\mathbf{h}_{\mathit{b}}[\mathit{k}]\parallel^{2}\right\}\geq \textrm{Tr}\left\{\mathbf{J}^{-1}[\mathit{k}]\right\} \\
& = \textrm{Tr}\left\{\left(\tilde{\mathbf{\Psi}}_{\mathit{p}}^{\mathit{H}}[\mathit{k}]\mathbf{R}_{\mathit{v}_{\mathit{p}}}^{-1}\tilde{\mathbf{\Psi}}_{\mathit{p}}[\mathit{k}] + \hat{\mathbf{\Gamma}}_{\mathit{k}}^{-1}\right)^{-1}\right\}. \label{estimate_men}
\end{split}
\end{align}
Upon exploiting the relationship between the beamspace representation of the THz MIMO channel and its vectorized version, the BCRLB for the estimated CSI is finally given by
\begin{align}
\textrm{MSE}(\hat{\mathbf{H}}[\mathit{k}]) \geq \textrm{Tr}\left\{\mathbf{\Psi}[\mathit{k}]\mathbf{J}^{-1}[\mathit{k}]\mathbf{\Psi}^{\mathit{H}}[\mathit{k}]\right\}. \label{bcrlb}
\end{align}
A drawback of THz systems is the fact that the received signal is severely affected by noise in the training phase. Therefore, it is difficult to obtain an accurate CSI estimate using exclusively pilot symbols \cite{alzeer2018millimeter}. In order to overcome this limitation, the DA-BL framework is proposed next which leverages the data along with the pilot symbols for enhancing the estimation accuracy.
\section{DA-BL For Sparse Channel Estimation in THz Hybrid MIMO systems}
The data-model is developed next for our THz MIMO system, which facilitates joint channel estimation and data-detection. Let $\mathbf{x}_{\mathit{d},m}[\mathit{k}]$ denote the data symbol vector whose elements are derived from an appropriate constellation. The data model corresponding to the $\mathit{k}$-th subcarrier is given by
\begin{align}
\mathbf{y}_{\mathit{d},m}[\mathit{k}] = \mathbf{W}_{\mathit{RF},m}^{\mathit{H}}\mathbf{H}[\mathit{k}]\mathbf{F}_{\mathit{RF},m}\mathbf{x}_{\mathit{d},m}[\mathit{k}] + \mathbf{W}_{\mathit{RF},m}^{\mathit{H}}\mathbf{v}_{\mathit{d},m}[\mathit{k}], \label{data_output}
\end{align}
where we have $\mathbf{y}_{d,m}[k] \in \mathbb{C}^{N_{RF} \times 1}$,$\mathbf{W}_{\mathit{RF},m}\in\mathbb{C}^{\mathit{N}_{\mathit{R}}\times\mathit{N}_{\mathit{RF}}}$ and $\mathbf{F}_{\mathit{RF},m}\in\mathbb{C}^{\mathit{N}_{\mathit{T}}\times\mathit{N}_{\mathit{RF}}}$. The construction of $\mathbf{W}_{\mathit{RF},m}$ and $\mathbf{F}_{\mathit{RF},m}$ is explained in Section VII-A. In \eqref{data_output} $\mathbf{x}_{\mathit{d},m}[\mathit{k}]\in\mathbb{C}^{\mathit{N}_{\mathit{RF}}\times 1}$ and $\mathbf{v}_{\mathit{d},m}[\mathit{k}] \in \mathbb{C}^{N_{R} \times 1}$ represents the $\mathit{K}$-point FFT of $\mathbf{v}_{\mathit{d},m}$, where $\mathbf{v}_{\mathit{d},m}\sim\mathcal{CN}(\mathbf{0}_{N_R \times 1},\sigma^2_d\mathbf{I}_{N_R})$. Let $\tilde{\mathbf{v}}_{\mathit{d},m}[\mathit{k}] = \mathbf{W}_{\mathit{RF},m}^{\mathit{H}}\mathbf{v}_{\mathit{d},m}[\mathit{k}]$ represent the equivalent noise after receiver combining and $\mathbf{H}_{\textrm{eq},m}[\mathit{k}] = \mathbf{W}_{\mathit{RF},m}^{\mathit{H}}\mathbf{H}[\mathit{k}]\mathbf{F}_{\mathit{RF},m}\in\mathbb{C}^{\mathit{N}_{\mathit{RF}}\times\mathit{N}_{\mathit{RF}}}$ represent the equivalent channel matrix. Using equation \eqref{data_output}, one can readily derive the expressions for the conventional ZF and MMSE based detectors, which are respectively given by
\begin{align}
\hat{\mathbf{x}}_{\mathit{d},m}^{\textrm{ZF}}[\mathit{k}] = \left(\mathbf{H}_{\textrm{eq},m}^{\mathit{H}}[\mathit{k}]\mathbf{H}_{\textrm{eq},m}[\mathit{k}]\right)^{-1}\mathbf{H}_{\textrm{eq},m}^{\mathit{H}}[\mathit{k}]\mathbf{y}_{\mathit{d},m}[\mathit{k}], \label{ls_data}
\end{align}
\begin{align}
\hat{\mathbf{x}}_{\mathit{d},m}^{\textrm{MMSE}}[\mathit{k}] = & \left(\mathbf{H}_{\textrm{eq},m}^{\mathit{H}}[\mathit{k}]\mathbf{H}_{\textrm{eq},m}[\mathit{k}] + \mathbf{R}_{\mathit{d},m}\right)^{-1}\mathbf{H}_{\textrm{eq},m}^{\mathit{H}}[\mathit{k}]\mathbf{y}_{\mathit{d},m}[\mathit{k}], \label{mmse_data}
\end{align}
where $\mathbf{R}_{\mathit{d},m}$ represents the noise covariance matrix for $\mathit{k}$-th subcarrier, which is defined as 
\begin{align}
\mathbf{R}_{d,m} = \mathbb{E}\left\{\tilde{\mathbf{v}}_{\mathit{d},m}[\mathit{k}]\tilde{\mathbf{v}}_{\mathit{d},m}^{\mathit{H}}[\mathit{k}]\right\} = \sigma^2_d \mathit{K}\mathbf{W}_{\mathit{RF},m}^{\mathit{H}}\mathbf{W}_{\mathit{RF},m}. \label{noise_data_cov}
\end{align}
The CSI estimate obtained in the previous section using the PA-BL algorithm is used for obtaining the initial data symbol estimate of the MMSE estimator. Note that during the estimation process, $\mathbf{W}^H_{RF, m}$ is determined as per Section VII-A, while $\mathbf{W}^H_{BB,m}[k]$ is set as the identity matrix. During the following data detection phase, $\mathbf{W}^H_{BB,m}[k]$ is designed as the LMMSE equalizer $\left(\mathbf{H}_{\textrm{eq},m}^{\mathit{H}}[\mathit{k}]\mathbf{H}_{\textrm{eq},m}[\mathit{k}]+ \mathbf{R}_{\mathit{d},m}\right)^{-1}\mathbf{H}_{\textrm{eq},m}^{\mathit{H}}[\mathit{k}]$. By expliting the characterstics of the $\textrm{vec}(.)$ operator as discussed in Section-I-C, one can recast the system model in equation \eqref{data_output} as
\vspace{-3pt}
\begin{align}
\mathbf{y}_{\mathit{d},m}[\mathit{k}] = \mathbf{\Phi}_{\mathit{d},m}[\mathit{k}]\mathbf{h}[\mathit{k}] + \tilde{\mathbf{v}}_{\mathit{d},m}[\mathit{k}], \label{vec_data_beam}
\end{align}
where the equivalent sensing matrix $\mathbf{\Phi}_{\mathit{d},m}[\mathit{k}]\in\mathbb{C}^{\mathit{N}_{\mathit{RF}}\times\mathit{N}_{\mathit{T}}\mathit{N}_{\mathit{R}}}$ is given by
\vspace{-5pt}
\begin{align}
\mathbf{\Phi}_{\mathit{d},m}[\mathit{k}] = (\mathbf{x}_{\mathit{d},m}^{\mathit{T}}[\mathit{k}]\mathbf{F}_{\mathit{RF},m}^{\mathit{T}}\otimes\mathbf{W}_{\mathit{RF},m}^{\mathit{H}}). \label{data_sens}
\end{align}
Finally, upon stacking the outputs $\mathbf{y}_{\mathit{d},m}[\mathit{k}]$ across all the data blocks, one obtains
\begin{align}
\underbrace{\begin{bmatrix}{\mathbf{y}}_{d,\text{1}}[k]\\ \vdots\\ {\mathbf{y}}_{d,N_d}[k]\end{bmatrix}}_{{\mathbf{y}_{\mathit{d}}}[k]} = \underbrace{ \begin{bmatrix} \mathbf{\Phi}_{d,\text{1}}[\mathit{k}]\\ \vdots\\ \mathbf{\Phi}_{d,N_d}[\mathit{k}]\end{bmatrix}}_{\mathbf{\Phi_{\mathit{d}}[\mathit{k}]}}\mathbf{h}[\mathit{k}] + \underbrace{\begin{bmatrix} \tilde{\mathbf{v}}_{d,\text{1}}[k]\\ \vdots\\ \tilde{\mathbf{v}}_{d,N_d}[k]\end{bmatrix}}_{\mathbf{v}_{\mathit{d}}[k]}, \label{data_con}
\end{align}
where $\mathbf{y}_d[k] \in \mathbb{C}^{N_d N_{RF} \times 1}$, $\mathbf{\Phi}_d[k] \in \mathbb{C}^{N_d N_{RF} \times N_T N_R}$ and $\mathbf{v}_d[k] \in \mathbb{C}^{N_d N_{RF} \times 1}$. Thus, the noise covariance matrix obeys $\mathbf{R}_{v_d} \in \mathbb{C}^{N_d N_{RF} \times N_d N_{RF}}$ where $\mathbf{R}_{v_d} = \sigma^2_d K \: \text{blkdiag}\left(\left\{\mathbf{W}_{RF,m}^H \mathbf{W}_{RF,m}\right\}_{m=1}^M\right)$. Eventually, to develop the joint CSI estimation model and data-detection, one can further concatenate the outputs corresponding to the data and pilot models i.e $\mathbf{y}_d[k]$ and $\mathbf{y}_p[k]$ respectively \cite{srivastava2021data}, from equation \eqref{data_con} and \eqref{con_frame} as
\begin{align}
\underbrace{\begin{bmatrix}{\mathbf{y}}_{\mathit{d}}[\mathit{k}]\\ {\mathbf{y}}_{\mathit{p}}[\mathit{k}]\end{bmatrix}}_{{\mathbf{y}}[\mathit{k}] \ \in \ \mathbb{C}^{\left(\mathit{N}_{\mathit{d}}+\mathit{M}\right)\mathit{N}_{\mathit{RF}}\times1}} = & \underbrace{ \begin{bmatrix} \mathbf{\Phi}_{\mathit{d}}[\mathit{k}]\\ \mathbf{\Phi}_{\mathit{p}}[\mathit{k}]\end{bmatrix}}_{\mathbf{\Phi[\mathit{k}] \ \in \ \mathbb{C}^{(\mathit{N}_{\mathit{d}}+\mathit{M})\mathit{N}_{\mathit{RF}}\times\mathit{N}_{\mathit{T}}\mathit{N}_{\mathit{R}}}}}\mathbf{h}[\mathit{k}] + \notag \\ & \underbrace{\begin{bmatrix} \mathbf{v}_{\mathit{d}}[\mathit{k}] \\ \mathbf{v}_{\mathit{p}}[\mathit{k}]\end{bmatrix}}_{\mathbf{v}[k] \ \in \ \mathbb{C}^{(\mathit{N}_{\mathit{d}}+\mathit{M})\mathit{N}_{\mathit{RF}}\times1}}. \label{conc_data_pilot}
\end{align}
Furthermore, the equivalent sensing matrix for the data-aided CSI estimation model is given by $\tilde{\mathbf{\Psi}}[\mathit{k}] = \mathbf{\Phi}[\mathit{k}]\mathbf{\Psi}[\mathit{k}]\in\mathbb{C}^{\left(\mathit{N}_{\mathit{d}}+\mathit{M}\right)\mathit{N}_{\mathit{RF}}\times\mathit{G}_{\mathit{R}}\mathit{G}_{\mathit{T}}}$. The covariance matrix $\mathbf{R}_{v}\in\mathbb{C}^{\left(\mathit{N}_{\mathit{d}}+\mathit{M}\right)\mathit{N}_{\mathit{RF}}\times\left(\mathit{N}_{\mathit{d}}+\mathit{M}\right)\mathit{N}_{\mathit{RF}}}$ of the noise vector $\mathbf{v}[\mathit{k}]$ is 
\vspace{-15pt}
\begin{align}
\mathbf{R}_{v} = \textrm{blkdiag}\left(\mathbf{R}_{{v}_{\mathit{d}}},\mathbf{R}_{{v}_{\mathit{p}}}\right). \label{conc_cov}
\end{align}
The proposed DA-BL scheme estimates the hyper-parameter matrix $\mathbf{\Gamma}_{\mathit{k}}$ using the EM procedure described below. Let the entire information set for this be defined as $\boldsymbol{\eta}_k = \left\{\mathbf{X}_{\mathit{d}}[\mathit{k}],\mathbf{\Gamma}_{\mathit{k}}\right\}$, where $\mathbf{X}_d[k] \in \mathbb{C}^{N_{RF} \times N_d}$ obtained by horizontally concatenating the transmit data vectors across all blocks and the parameter estimate at iteration $(j-1)$ be given as
\begin{align}
\hat{\boldsymbol{\eta}}_k^{\left(\mathit{j}-1\right)} = \left\{\hat{\mathbf{X}}_{\mathit{d}}^{\left(\mathit{j}-1\right)}[\mathit{k}],\hat{\mathbf{\Gamma}}_{\mathit{k}}^{\left(\mathit{j}-1\right)}\right\}, \label{par_eqn}
\end{align}
where $\hat{\mathbf{X}}_{\mathit{d}}^{\left(\mathit{j}-1\right)}[\mathit{k}]$ denotes the estimate of the data in the $(\mathit{j} - 1)$st iteration, whereas $\hat{\mathbf{\Gamma}}_{\mathit{k}}^{\left(\mathit{j}-1\right)}$ denotes the hyper-parameter matrix obtained in the $(\mathit{j}-1)$st iteration. The conditional expectation of the log-likelihood function $\mathcal{L}(\boldsymbol{\eta}_k|\hat{\boldsymbol{\eta}_k}^{\left(\mathit{j}-1\right)})$ calculated in the E-step for the $j$th iteration is given by
\begin{align}
\mathcal{L}(\boldsymbol{\eta}_k|\hat{\boldsymbol{\eta}_k}^{\left(\mathit{j}-1\right)}) &= \mathbb{E}_{\mathbf{h}_{\mathit{b}}[\mathit{k}]|\mathbf{y}[\mathit{k}];\hat{\boldsymbol{\eta}_k}^{\left(\mathit{j}-1\right)}}\left\{\textrm{log}\left[\mathit{p}\left(\mathbf{y}[\mathit{k}],
\mathbf{h}_{\mathit{b}}[\mathit{k}];\boldsymbol{\eta}_k\right)\right]\right\} \notag \\
&= \mathbb{E}\left\{\textrm{log}\left[\mathit{p}\left(\mathbf{y}[\mathit{k}]|\mathbf{h}_{\mathit{b}}[\mathit{k}];\mathbf{X}_{\mathit{d}}[\mathit{k}]\right)\right]\right\} + \notag \\ & \quad\quad\quad\quad \mathbb{E}\left\{\textrm{log}\left[\mathit{p}\left(\mathbf{h}_{\mathit{b}}[\mathit{k}];\mathbf{\Gamma}_{\mathit{k}}\right)\right]\right\}. \label{log_data}
\end{align}
By simplifying the first term in the above expectation, we obtain
\vspace{-5pt}
\begin{align}
   &\mathcal{L}(\mathit{p}\left(\mathbf{y}[\mathit{k}]|\mathbf{h}_{\mathit{b}}[\mathit{k}];\mathbf{X}_{\mathit{d}}[\mathit{k}]\right))= -(\mathit{N}_{\mathit{d}}+\mathit{M})\mathit{N}_{\mathit{RF}}\textrm{log}(\pi) - \notag \\ & \textrm{log}\left[\textrm{det}(\mathbf{R}_{v})\right] - \left(\mathbf{y}[\mathit{k}]-\tilde{\mathbf{\Psi}}[\mathit{k}]\mathbf{h}_{\mathit{b}}[\mathit{k}]\right)^{\mathit{H}}\mathbf{R}_{v}^{-1} \times \notag \\ & \quad\quad\quad\quad\quad\quad\quad\quad\quad\quad\quad\quad \left(\mathbf{y}[\mathit{k}]-\tilde{\mathbf{\Psi}}[\mathit{k}]\mathbf{h}_{\mathit{b}}[\mathit{k}]\right). \label{con_data_aided}
\end{align}
\begin{algorithm}[t!]
\caption{BL-based channel estimation and robust data detection for THz MIMO systems}\label{algo_DataAided}
\textbf{Input:} Observation vector $\mathbf{y}[\mathit{k}]$, Data output $\mathbf{Y}_{\mathit{d}}[\mathit{k}]$, Pilot sensing matrix $\mathbf{\Phi}_{\mathit{p}}[\mathit{k}]$, Equivalent sensing matrix $\tilde{\mathbf{\Psi}}[\mathit{k}]$, Noise Covariance $\mathbf{R}_{\mathit{v}}$, dictionary matrices $\mathbf{A}_{\mathit{R}}(\Theta_{\mathit{R}},\mathit{f}_{\mathit{k}})$ and $\mathbf{A}_{\mathit{T}}(\Theta_{\mathit{T}},\mathit{f}_{\mathit{k}})$, stopping parameters $\epsilon$ and $\mathit{J}_{\textrm{max}}$\\
\textbf{Initialization:} $\hat{\mathbf{X}}_{\mathit{d}}^{(\textrm{0})}[\mathit{k}] = \hat{\mathbf{X}}_{\mathit{d},\mathit{p}}^{\textrm{MMSE}}[\mathit{k}]$ estimated using the CSI obtained from PA-BL algorithm, counter $\mathit{j} = 0$ \\
\textbf{while}$(\parallel\hat{\mathbf{\Gamma}}_{\mathit{k}}^{\left(\mathit{j}\right)} - \hat{\mathbf{\Gamma}}_{\mathit{k}}^{\left(\mathit{j}-1\right)}\parallel_\mathit{F} > \epsilon)$ and $(\mathit{j}<\mathit{J}_{\textrm{max}})$ \textbf{do}
\begin{itemize}
\item[]$\mathit{j} \leftarrow \mathit{j}+1$
\item[] \textbf{E-step:} Compute the \textit{aposteriori} covariance and mean
\begin{equation}
\boldsymbol{\Sigma}^{\left(\mathit{j}\right)}[\mathit{k}] =\left[(\tilde{\mathbf{\Psi}}^{(j-1)}[\mathit{k}])^H\mathbf{R}_{\mathit{v}}^{-\textrm{1}}\tilde{\mathbf{\Psi}}^{(j-1)}[\mathit{k}] + \left(\hat{\mathbf{\Gamma}}_{\mathit{k}}^{\left(\mathit{j} - 1\right)}\right)^{-\textrm{1}}\right]^{-\textrm{1}}\notag
\end{equation}
\begin{equation}
\hat{\mathbf{h}}_{\mathit{b}}^{\left(\mathit{j}\right)}[\mathit{k}] = \boldsymbol{\Sigma}^{\left(\mathit{j}\right)}[\mathit{k}](\tilde{\mathbf{\Psi}}^{\left(\mathit{j}-1\right)}[\mathit{k}])^{\mathit{H}}\mathbf{R}_{\mathit{v}}^{-\textrm{1}}\mathbf{y}[\mathit{k}]\notag
\end{equation}
\item[] \textbf{M-step1:} Update the hyperparameters
    \begin{equation}
    \begin{split}
\hat{\gamma}_{\mathit{k},\mathit{i}}^{\left(\mathit{j}\right)} = \boldsymbol{\Sigma}^{\left(\mathit{j}\right)}[\mathit{k}](\mathit{i},\mathit{i}) + |\hat{\mathbf{h}}_{\mathit{b}}^{\left(\mathit{j}\right)}[\mathit{k}](\mathit{i})|^{2};\quad
1\leq\mathit{i}\leq\mathit{G}_{
\mathit{R}}\mathit{G}_{\mathit{T}}\notag
\end{split}
\end{equation}
\item[] \textbf{M-step2:}
\begin{enumerate}
\item Evaluate $\overline{\mathbf{Y}}_{d}[k]$ and $\hat{\boldsymbol{\mathcal{H}}}_{eq}^{(j)}[k]$ using equation \eqref{cal_of_updated_esti}
\item Update the data estimate $\hat{\mathbf{X}}_{d}^{\left(j\right)}[k]$ using \eqref{zf_vec_data}
\item Demodulate the estimated data and use it to update $\tilde{\mathbf{\Psi}}^{\left(\mathit{j}\right)}[\mathit{k}]$ 
\end{enumerate}
\end{itemize}
\textbf{end while}\\
$\hat{\mathbf{h}}_{\mathit{b}}[\mathit{k}] = \hat{\mathbf{h}}_{\mathit{b}}^{\left(\mathit{j}\right)}[\mathit{k}]$\\
\textbf{Output:}  $\hat{\mathbf{H}}_{\textrm{DA-BL}}[\mathit{k}] = \mathbf{A}_{\mathit{R}}(\Theta_{\mathit{R}},\mathit{f}_{\mathit{k}})\textrm{vec}^{-1}(\hat{\mathbf{h}}_{\mathit{b}}[\mathit{k}])\mathbf{A}_{\mathit{T}}^{\mathit{H}}(\Theta_{\mathit{T}},\mathit{f}_{\mathit{k}})$, \quad
$\hat{\mathbf{X}}_{\mathit{d}}[\mathit{k}] = \hat{\mathbf{X}}_{\mathit{d}}^{\left(\mathit{j}\right)}[\mathit{k}]$ \\
\end{algorithm}
Observe that the initial term relies only on the data matrix $\mathbf{X}_{d}[k]$ through the matrix $\tilde{\mathbf{\Psi}}[\mathit{k}]$. In a similar fashion, one can see that the second term is solely determined by the hyper-parameter matrix $\boldsymbol{\Gamma}_{\mathit{k}}$
\begin{align}
    \mathcal{L}(p\left[\mathbf{h}_b[k];\mathbf{\Gamma}_k\right]) = -G_R G_T \mathrm{log}(\pi) - & \mathrm{log}[\mathrm{det}(\mathbf{\Gamma}_k)]- \notag \\ & \mathbf{h}^H_b[k]\mathbf{\Gamma}^{-1}_k\mathbf{h}_b[k]. \label{FIM_beam_data} 
\end{align}
Thus, the joint maximization of the log-likelihood $\mathcal{L}(\boldsymbol{\eta}_k|\hat{\boldsymbol{\eta}_k}^{\left(\mathit{j}-1\right)})$ with respect to $\boldsymbol{\eta}_k$ in the M-step breaks down to two independent maximization steps, one with respect to $\mathbf{X}_{d}[k]$ and the other with $\boldsymbol{\Gamma}_{\mathit{k}}$. The optimization problem in the M-step corresponding to the hyper-parameter update is given by
\begin{align}
\hat{\gamma}_{\mathit{k},\mathit{i}}^{\left(\mathit{j}\right)} = \underset{\gamma_{\mathit{k,i}}\geq0}{\textrm{arg max}} \ &  \mathbb{E}_{\mathbf{h}_{\mathit{b}}[\mathit{k}]|\mathbf{y}[\mathit{k}];\hat{\boldsymbol{\eta}_k}^{\left(\mathit{j}-1\right)}}\left\{\textrm{log}\left[\mathit{p}\left(\mathbf{h}_{\mathit{b}}[\mathit{k}];\boldsymbol{\Gamma}_{\mathit{k}}\right)\right]\right\} \notag \\ & \quad\quad\quad\quad\quad\quad\quad\quad
\forall \ 1\leq\mathit{i}\leq\mathit{G}_{\mathit{R}}\mathit{G}_{\mathit{T}}. \label{mstep_dabl}
\end{align}
Along similar lines to the PA-BL technique, the estimate of $\gamma_{\mathit{k},\mathit{i}}$ in iteration $j$ is given by
\vspace{-3pt}
\begin{align}
\hat{\gamma}_{\mathit{k},\mathit{i}}^{\left(\mathit{j}\right)} & = \mathbb{E}_{\mathbf{h}_{\mathit{b}}[\mathit{k}]|\mathbf{y}[\mathit{k}];\hat{\boldsymbol{\Gamma}}_k^{\left(\mathit{j-\textrm{1}}\right)}}\left\{|\mathbf{h}_{\mathit{b}}[\mathit{k}](\mathit{i})|^{2}\right\} \notag \\ & =  \boldsymbol{\Sigma}^{\left(\mathit{j}\right)}[\mathit{k}](\mathit{i},\mathit{i}) + |\hat{\mathbf{h}}_{\mathit{b}}^{\left(\mathit{j}\right)}[\mathit{k}](\mathit{i})|^{2}, \label{sol_mstep_dabl}
\end{align}
where the \textit{a posteriori} mean vector $\hat{\mathbf{h}}_{\mathit{b}}^{\left(\mathit{j}\right)}[\mathit{k}] \in \mathbb{C}^{G_R G_T \times1}$ is formulated as
\vspace{-3pt}
\begin{align}
\hat{\mathbf{h}}_{\mathit{b}}^{\left(\mathit{j}\right)}[\mathit{k}] = \boldsymbol{\Sigma}^{\left(\mathit{j}\right)}[\mathit{k}]\left(\tilde{\boldsymbol{\Psi}}^{\left(\mathit{j-\textrm{1}}\right)}[\mathit{k}]\right)^{\mathit{H}}\mathbf{R}_{v}^{-1}\mathbf{y}[\mathit{k}], \label{mean_dabl}
\end{align}
and its covariance matrix $\boldsymbol{\Sigma}^{\left(\mathit{j}\right)}[\mathit{k}] \in \mathbb{C}^{G_R G_T \times G_R G_T}$ is given by the expression
\vspace{-5pt}
\begin{align}
\boldsymbol{\Sigma}^{\left(\mathit{j}\right)}[\mathit{k}] =\left[(\tilde{\mathbf{\Psi}}^{(j-1)}[\mathit{k}])^H\mathbf{R}_{\mathit{v}}^{-\textrm{1}}\tilde{\mathbf{\Psi}}^{(j-1)}[\mathit{k}] + \left(\hat{\mathbf{\Gamma}}_{\mathit{k}}^{\left(\mathit{j} - 1\right)}\right)^{-\textrm{1}}\right]^{-\textrm{1}}. \label{cov_dabl}
\end{align}
The matrix $\tilde{\mathbf{\Psi}}^{(j-\textrm{1})}[\mathit{k}]$ that appears in the above expression depends on the data symbol matrix $\hat{\mathbf{X}}_{d}^{(j-1)}[k]$ in the $\left(j-\textrm{1}\right)$st iteration. Therefore, the M-step of the optimization procedure toward determining the data matrix update $\hat{\mathbf{X}}_d^{\left(j\right)}[k]$ can be formulated as
\vspace{-7pt}
\begin{align}
\hat{\mathbf{X}}_{d}^{(j)}[k] =  \arg \max_{ \mathbf{X}_{d}[k] } \mathbb{E} \bigg\{  \log \left[ p\left(\mathbf{y}_{d}[k]\mid \mathbf{h}_{b}[k]; \mathbf{X}_{d}[k] \right) \right] \bigg\}. \label{M_step2}
\end{align}
Using equation \eqref{data_output}, the above maximization objective can be reformulated as seen in equation \eqref{eq:decouple_max_3}. This can be further simplified as seen in \eqref{eq:decouple_max_4}-\eqref{eq:decouple_max_6}. After further simplification, \eqref{M_step2} can be reframed as
\vspace{-6pt}
\begin{align}
\hat{\mathbf{X}}_{d}^{\left(j\right)}[k]    =& \arg \min_{\substack{\mathbf{X}_{d}[k]}}\left\lbrace \left\|\overline{\mathbf{Y}}_{d}[k] -\widehat{\boldsymbol{\mathcal{H}}}_{eq}^{(j)}[k]\mathbf{X}_{d}[k]\right\|_F^2 \right\rbrace \label{eq:decouple_max_7},
\end{align}
where, we have
\vspace{-5pt}
\begin{align}
\overline{\mathbf{Y}}_{d}[k]=\begin{bmatrix} 
      \mathbf{Y}_{d}[k] \\
     \mathbf{0}\\
       \end{bmatrix}, \ \  \hat{\boldsymbol{\mathcal{H}}}_{eq}^{(j)}[k]=\begin{bmatrix}
       \hat{\mathbf{H}}_{\textrm{eq}}^{(j)}[k] \\ 
      \left(\boldsymbol{\Xi}^{(j)}[k]\right)^{\frac{1}{2}}\\
       \end{bmatrix}, \label{cal_of_updated_esti}
\end{align}
and $\mathbf{Y}_d[k] \in \mathbb{C}^{N_{RF} \times N_d}$ is obtained by horizontally concatenating the received data vectors across all blocks. The quantity 
\begin{figure*}
 \begin{align}
    \hat{\mathbf{X}}_{d}^{\left(j\right)}[k]=& \arg \max_{\substack{\mathbf{X}_{d}[k]}} \mathbb{E}_{\mathbf{H}_{\textrm{eq}}[k]\mid \mathbf{Y}_{d}[k];\hat{\mathbf{X}}_{d}^{\left(j-1\right)}[k]} \Bigg\{ \log \left[ p\left(\mathbf{Y}_{d}[k]\mid \mathbf{H}_{\textrm{eq}}[k]; \mathbf{X}_{d}[k]\right) \right]\Bigg\} \label{eq:decouple_max_3}\\
     =&\arg \min_{\substack{\mathbf{X}_{d}[k]}} \mathbb{E}_{\mathbf{H}_{\textrm{eq}}[k]\mid \mathbf{Y}_{d}[k];\hat{\mathbf{X}}_{d}^{\left(j-1\right)}[k]} \Bigg\{ \left\Vert \mathbf{Y}_{d}[k]-\mathbf{H}_{\textrm{eq}}[k]\mathbf{X}_{d}[k]\right\Vert_{F}^2 \Bigg\}  \label{eq:decouple_max_4} \\
    =&\arg \min_{\substack{\mathbf{X}_{d}[k]}} \left\{\left\|\mathbf{Y}_{d}[k]-\hat{\mathbf{H}}_{\textrm{eq}}^{(j)}[k]\mathbf{X}_{d}[k]\right\|_F^2+\left\|\left(\boldsymbol{\Xi}^{(j)}[k]\right)^{\frac{1}{2}} \mathbf{X}_{d}[k] \right\|_F^2\right\} \label{eq:decouple_max_5}\\
    =&\arg \min_{\substack{\mathbf{X}_{d}[k]}}\left\lbrace \left \|\begin{bmatrix} 
    \hspace{-10pt}\mathbf{Y}_{d}[k] \\
    \mathbf{0}_{N_{RF}\times N_d}\\
      \end{bmatrix} -\begin{bmatrix} \hat{\mathbf{H}}_{\textrm{eq}}^{(j)}[k] \\ \left(\boldsymbol{\Xi}^{(j)}[k]\right)^{\frac{1}{2}}\\
      \end{bmatrix}\mathbf{X}_{d}[k]\right\|_F^2 \right\rbrace  \label{eq:decouple_max_6}
     \end{align}
\hrulefill
\end{figure*}
$\hat{\mathbf{H}}_{\textrm{eq}}^{\left(\mathit{j}\right)}[\mathit{k}] \in \mathbb{C}^{\mathit{N}_{\mathit{RF}}\times\mathit{N}_{\mathit{RF}}}$ above is defined as
\vspace{-3pt}
\begin{align}
\hat{\mathbf{H}}_{\textrm{eq}}^{\left(\mathit{j}\right)}[\mathit{k}] &= \mathbb{E} \left\{\mathbf{W}_{RF}^H \mathbf{H}[k] \mathbf{F}_{RF} \right\} \notag \\ & =\mathbf{W}_{\mathit{RF}}^{\mathit{H}}\mathbf{A}_{\mathit{R}}(\Theta_{\mathit{R}},\mathit{f}_{\mathit{k}})\textrm{vec}^{-1}\left(\hat{\mathbf{h}}_{\mathit{b}}^{\left(\mathit{j}\right)}[\mathit{k}]\right)
\mathbf{A}_{\mathit{T}}^{\mathit{H}}(\Theta_{\mathit{T}},\mathit{f}_{\mathit{k}})\mathbf{F}_{\mathit{RF}}, \label{eq_ch_info}
\end{align}
and the procedure of dividing $\boldsymbol{\Xi}^{(j)}[k]$ is detailed in Appendix-A. The ZF based detector used for updating the data symbol matrix is obtained as,
\vspace{-5pt}
\begin{align}
\hat{\mathbf{X}}_d^{\mathrm{ZF}(j)}[k] = \left\{\left(\hat{\boldsymbol{\mathcal{H}}}_{\mathrm{eq}}^{(j)}[k]\right)^{\mathit{H}}\hat{\boldsymbol{\mathcal{H}}}_{\mathrm{eq}}^{(j)}[k]\right\}^{-1}\left(\hat{\boldsymbol{\mathcal{H}}}_{\mathrm{eq}}^{(j)}[k]\right)^{\mathit{H}}\overline{\mathbf{Y}}_{d}[k]. \label{zf_vec_data}
\end{align}
The nearest neighbour detection rule is subsequently used to map each element in the matrix above to a symbol in the transmit constellation. The E-step and the M-step described above are executed iteratively until the algorithm converges. The various steps of the DA-BL framework are summarized in Algorithm $2$. The BCRLB for DA-BL framework is derived in Appendix B. In general, for a sparse signal recovery problem, when considering an over-complete dictionary, the performance of the proposed Bayesian learning schemes such as PA-BL and DA-BL seem to remain consistent even with a reduced number of observations. By contrast, the performance of non-Bayesian learning techniques such as Basis-Pursuit, FOCUSS, OMP degrade significantly, as observed in our simulation results.
\subsection{Computational Complexity and Memory Occupancy}
This subsection derives the computational complexity of the proposed PA-BL and DA-BL techniques for THz hybrid MIMO systems. Due to space constraints, the comprehensive calculations of the computational complexities for both the proposed and existing THz hybrid MIMO channel estimation methods are provided in our technical report \cite{abhisha}. The key findings from Tables I to III of the technical report are discussed next. It is evident from Table I that the OMP technique exhibits a worst-case complexity order of $\mathcal{O}(M^3 N_{RF}^3)$, attributed to the necessity of an intermediate least squares (LS) estimate at each iteration. As shown in Table II, the computational complexity of the PA-BL technique is of the order $\mathcal{O}(M N_{RF}^3 + G_R^3 G_T^3)$, which arises due to the matrix inversion of size $[G_RG_T \times G_RG_T]$ for computing the a posteriori covariance $\mathbf{R}_b^{(j)}[k]$. On the other hand, from Table III, one can see that the computational complexity of the proposed DA-BL technique is of the order $\mathcal{O}\big((N_d + M)^3 N_{RF}^3 + G_R^3 G_T^3\big)$, which is higher than that of the PA-BL and OMP-based techniques. However, the performance, as shown in the simulation results of Section VII, indicates superior performance for DA-BL. The OMP technique has the lowest complexity, but its performance is also poor. Furthermore, due to space constraints, we have similarly derived the memory occupancy for different channel estimation schemes, including OMP, PA-BL, and DA-BL in Table IV of our technical report \cite{abhisha}. It is evident from the memory and computational complexity analysis, that the proposed DA-BL technique has the highest memory requirement and computational complexity, but it also exhibits superior performance. Hence, one can see a clear trade-off performance vs. computational cost/ memory between the proposed and existing CSI estimation schemes.
\section{Simulation Results}
This section provides the performance evaluation of the proposed PA-BL and DA-BL techniques conceived for THz MIMO systems in the face of the dual-wideband effect. The amplitudes of the path coefficients corresponding to the LoS and NLoS elements are evaluated using equations \eqref{los_path} and \eqref{Nlos_path}, respectively, while the corresponding phase shifts are produced using a probability distribution that is uniform over the range (-$\pi$,$\pi$]. The coefficient of molecular absorption is calculated using equation \eqref{Lspread}, and its parameters have been taken from the HITRAN database \cite{hill}. The THz MIMO channel incorporating the dual-wideband effect is generated using $N_{cl} = 4$ clusters, namely one LoS and $\mathit{N}_{\textrm{NLoS}} = 3$ clusters. The carrier frequency utilized is $f_{c} = 1$ THz and the transmission distance considered is $d$ = 10 metres. We consider an indoor office scenario, where the molecular composition of air is given as water vapour $1\%$, oxygen $20.9\%$ and nitrogen $78.1\%$. The system temperature and pressure, respectively, are considered to be 1 atm and 296K for calculating the molecular absorption loss using equation \eqref{Lspread}. The AoAs/ AoDs associated with each multipath component have been generated from a Laplacian distribution, where the multipath component's angle exhibits a standard deviation of $\frac{1}{10}$ radian from the mean, which in turn is picked with uniform probability from an angular grid. Three reflecting media have been considered with the standard deviations of their roughness set as $\sigma \in {\left\{0.05,0.13,0.15\right\}}$ mm \cite{piesiewicz2007scattering}. The transmit and receive antenna gains are set to $A_T = 31$ dB, $A_R = 31$ dB, respectively.
Two THz MIMO systems are considered for simulations and their parameters are given in Table \ref{table2}. The grid sizes are chosen so that $G_T, G_R \geq $ max$(N_T,N_R)$. Both the systems have the number of pilot blocks $M$ set as $M \in \left\{30,50,70,90\right\}$. The spacings at the transmit and receive antenna ends are taken to be $d_t = d_r = \lambda/2$. Moreover, we consider the utilization of a raised cosine pulse shaping filter with a roll-off factor of 0.80 and an up-sampling factor of 20. The constituent elements of the transmitted pilot symbol vector $\mathbf{u}_{m,n}^{(p)}$ are chosen at unity average power from a 8-PSK constellation.
\subsection{Wideband THz hybrid MIMO channel estimation}
\begin{figure*}
	\centering
	\subfloat[]{\includegraphics[scale=0.32]{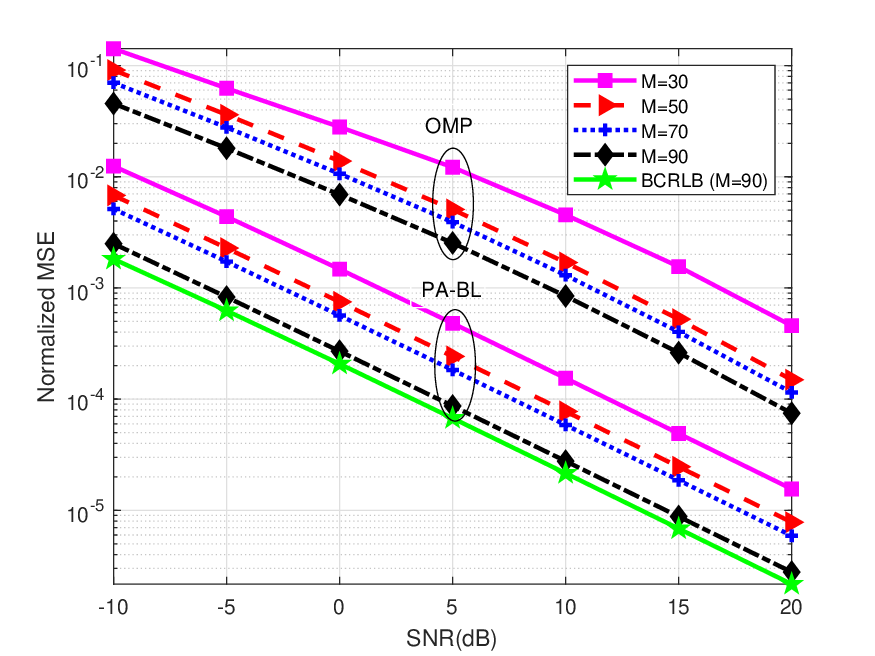}}
	\hfil
	\hspace{-10pt}\subfloat[]{\includegraphics[scale=0.32]{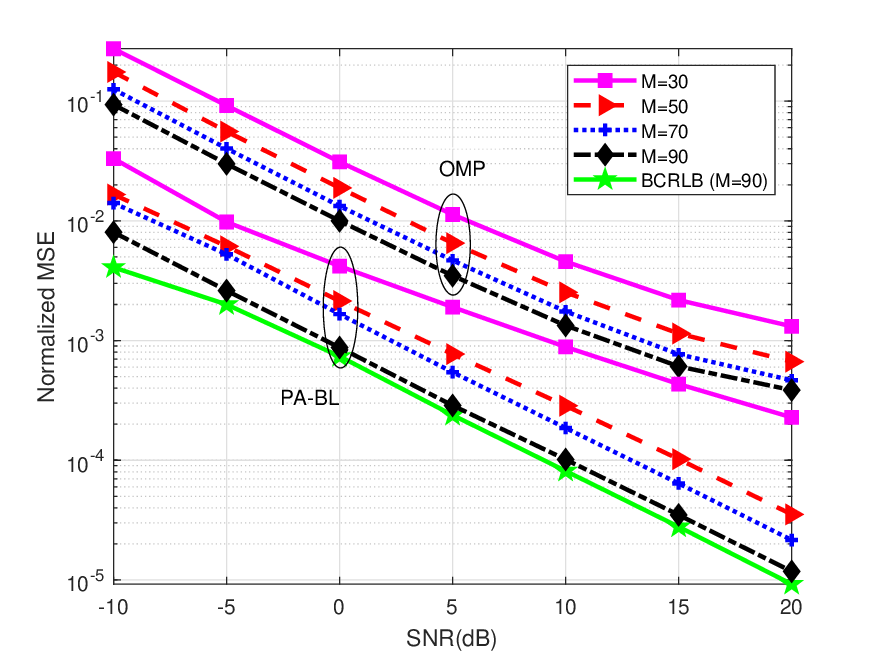}}
 	\hfil
	\hspace{-10pt} \subfloat[]{\includegraphics[scale=0.32]{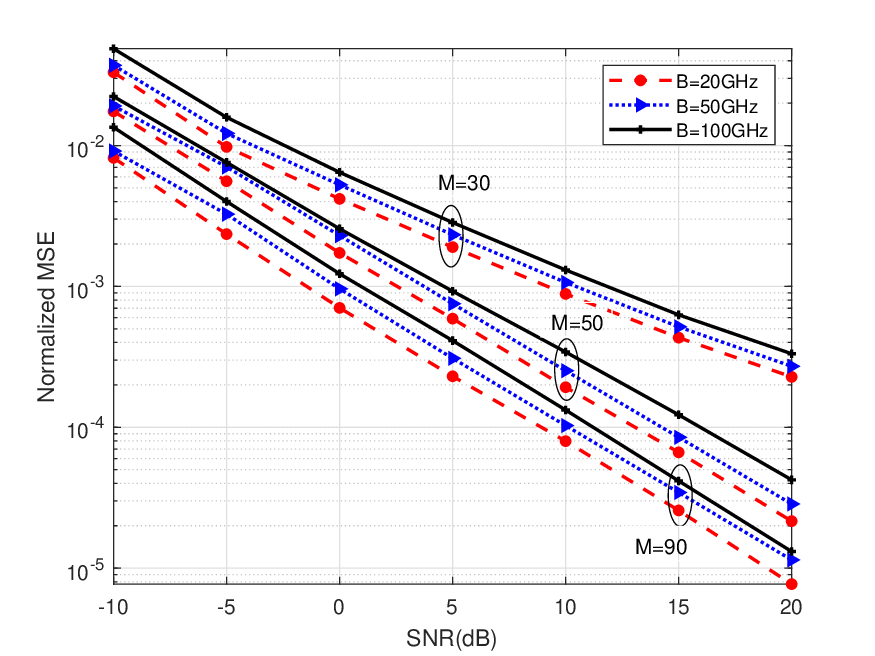}}
	\hfil
	\hspace{-10pt}\subfloat[]{\includegraphics[scale=0.32]{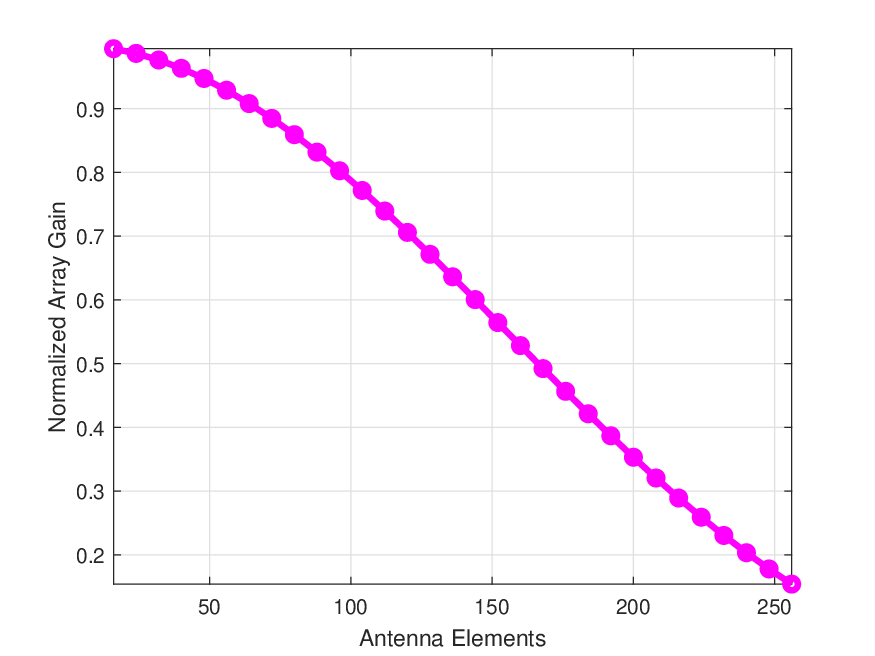}}
 \vspace{-5pt}
	\caption{Performance of the proposed PA-BL technique with the existing OMP technique for $ \left(a\right) $System-I $ \left(b\right) $System-II. $ \left(c\right) $ NMSE performance of the proposed PA-BL algorithm versus SNR, for various values of the bandwidth $B \in \left\{20,50,100\right\}$ GHz, $ \left(d\right) $ Normalized array gain versus number of antenna elements with bandwidth $B = 20$GHz at $f_c = 1$ THz.}
	\label{fig:DA_BL_iteration}
\end{figure*}

\begin{figure*} [t!]
	\centering
	\subfloat[]{\includegraphics[scale=0.42]{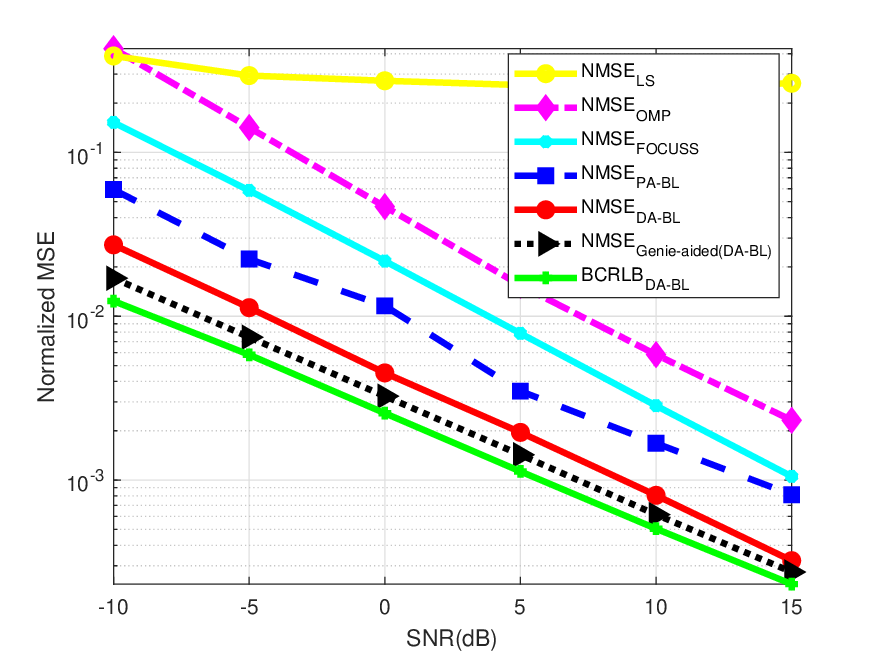}}
	\hfil
	\hspace{-10pt}\subfloat[]{\includegraphics[scale=0.42]{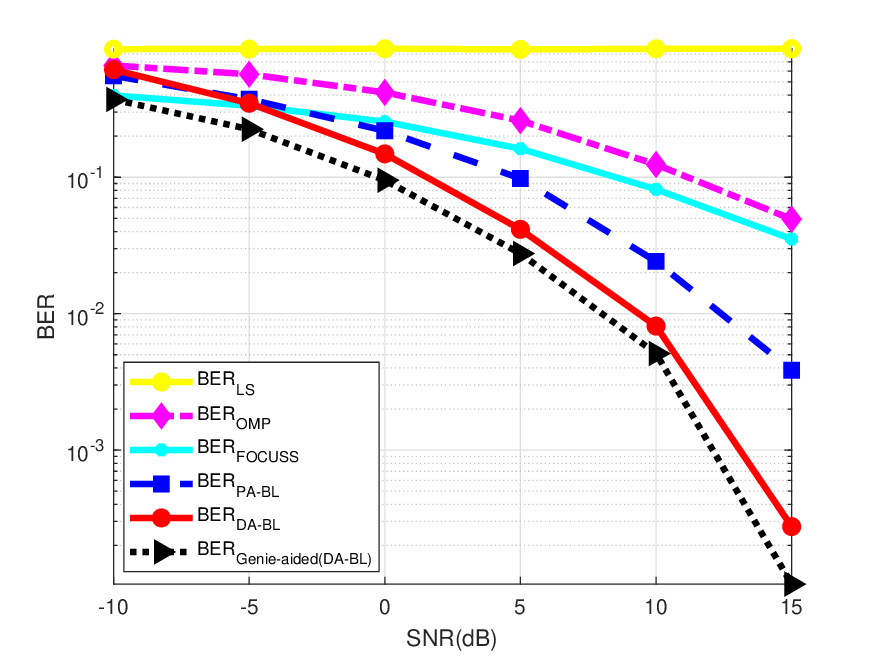}}
 	\hfil
	\hspace{-10pt}\subfloat[]{\includegraphics[scale=0.42]{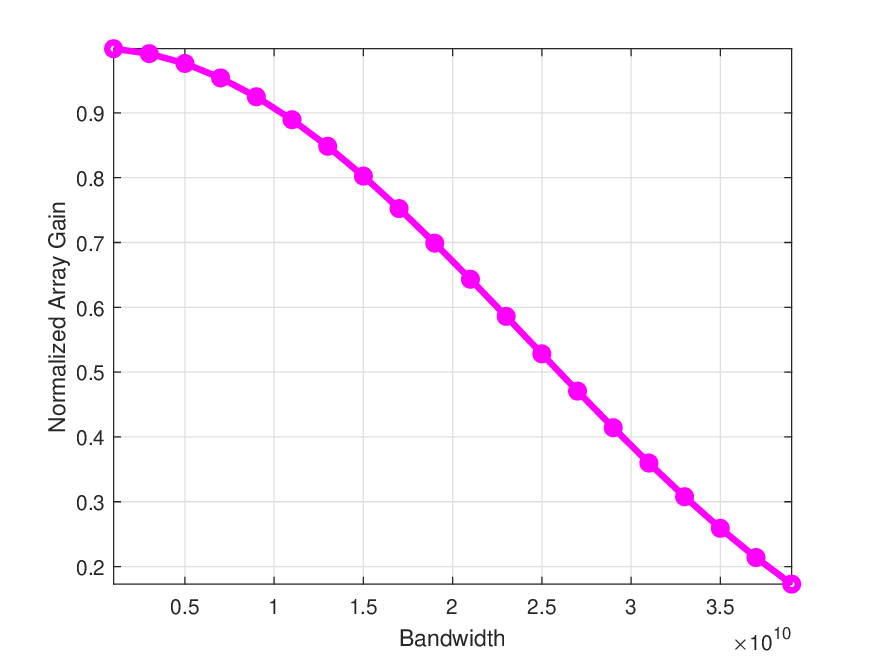}}
	  \caption{ $ \left(a\right) $ NMSE versus SNR comparison of the proposed PA-BL and DA-BL techniques with the Genie-aided estimator, LS, OMP, FOCUSS and the BCRLB, $ \left(b\right) $ BER versus SNR comparison of the proposed PA-BL and DA-BL techniques with the Genie-aided estimator for System-I. $\left(c\right)$ Normalized array gain versus bandwidth with $N_T = 128$ and $N_R = 128$ at $f_c = 1$THz.}
	\label{sys1}
\end{figure*}

The RF TPC and RC components, which are frequency-independent as shown in equation \eqref{frame_output}, are generated as \cite{srivastava2021fast}
\begin{align}
\mathbf{F}_{RF,m}\left(i,j\right) = \frac{1}{\sqrt{N_{t}}}\textrm{exp}\left(j\varphi_{i,j}\right), \notag
\end{align}
\vspace{-10pt}
\begin{align}
\mathbf{W}_{RF,m}\left(i,j\right) = \frac{1}{\sqrt{N_{r}}}\textrm{exp}\left(j\psi_{i,j}\right). \label{frf_wrf}
\end{align}
The phases $\varphi_{i,j}$ ,$\psi_{i,j}$ for the set $(i,j)$ are selected randomly with a uniform probability distribution across the set $\mathcal{A} =\left\{\textrm{0},\frac{2\pi}{2^{N_{Q}}},\cdots,\frac{\left(2^{N_{Q}}-1\right)2\pi}{2^{N_{Q}}}\right\}$ with $N_Q$ = 4 denoting the angle quantization parameter.
Fig. \ref{fig:DA_BL_iteration}(a) and (b) compares the normalized MSE (NMSE) of the channel estimates for the proposed PA-BL technique and the OMP technique for a dual-wideband THz MIMO channel, with parameters as per for Systems I and II. The tuning parameter for the OMP \cite{kulkarni2017low} algorithm which impacts the convergence of the algorithm is set as $\epsilon_o = 1$ for both the systems, while the stopping criterion and the upper limit on the number of PA-BL iterations are defined as $\epsilon = 1$ and $K_{max} = 20$, respectively. The NMSE metric is defined as $\textrm{NMSE} = \frac{\sum_{\mathit{k} = 0}^{\mathit{K}-1}\parallel\hat{\mathbf{H}}[k]-\mathbf{H}[k]\parallel_{F}^{2}}{\sum_{\mathit{k} = 0}^{\mathit{K}-1}\parallel\mathbf{H}[k]\parallel_{F}^{\textrm{2}}}.$ Observe from Fig. \ref{fig:DA_BL_iteration}(a) and (b), that the NMSE performance improves significantly as one increases the number of training blocks $M$. The explanation for this lies in the use of greater number of pilots which leads to more reliable CSI estimates. It is evident from the plot that the proposed PA-BL technique yields a considerably improved performance over the OMP algorithm for the entire SNR range under consideration. The existing OMP approach performs poorly owing to its susceptibility to changes in the stopping criterion and dictionary matrix. As a consequence, even slight modifications to the cessation criterion or deviations from the dictionary matrix constraints can result in severe structural and convergence problems.
\begin{table}[t]
\vspace{-8pt}
\centering
\caption{Simulation parameters for the dual-wideband THz system}
\label{table2}
\resizebox{0.42\textwidth}{!}{%
\begin{tabular}{|l|l|l|}\hline
\textbf{Parameter} & \textbf{System-I} & \textbf{System-II} \\ \hline \hline
$\#$ of TAs ($N_T$) & 32 & 64 \\\hline
$\#$ of RAs ($N_R$) & 32 & 64 \\ \hline
$\#$ of RF chains ($N_{RF}$) & 6 & 12 \\ \hline
$\#$ of diffused ray ($N_{ray}$) & 1 & 1 \\ \hline
$\#$ of subcarriers ($K$) & 16 & 32 \\ \hline
$\#$ of data vectors ($N_d$) & 100 & 200 \\ \hline
Transmit angular grid size & 64 & 128 \\ \hline
Receive angular grid size & 64 & 128 \\ \hline
\end{tabular}%
}
\vspace{-4pt}
\end{table}
The OMP approach also suffers from error propagation, because errors in the index selection cannot be corrected in subsequent iterations. It can be seen that the NMSE of the proposed PA-BL scheme improves by approximately $\approx$ 30 dB, when the SNR is varied from -10 dB to 20 dB. Upon increasing the number of pilot blocks $M$, the performance of the PA-BL scheme improves considerably. Interestingly, for a higher number of training frames such as $M = 90$, the proposed PA-BL algorithm performs close to the corresponding BCRLB. This is of crucial importance, as the BCRLB is calculated under the assumption of perfect knowledge of the AoAs/AoDs, which is an ideal case. In stark contrast, the PA-BL technique does not require any such prior information. Further, as demonstrated in \cite{wipf2004sparse}, upon convergence of the PA-BL algorithm, only the hyperparameters corresponding to the locations of the active AoA/ AoD pairs retain significant values, while the rest of the hyperparameters are driven to zero. By contrast, the performance of OMP and many other sparse recovery algorithms depends highly on the cessation criteria, which may be different for different propagation scenarios, thus leading to performance degradation, if not set carefully. Hence the proposed PA-BL algorithm is capable of learning the propagation environment in different network deployments. Fig. \ref{fig:DA_BL_iteration}(c), shows the performance of the PA-BL technique for increasing signal bandwidth $B$. As discussed in Section III-C, when the user bandwidth increases, the impact of the beam squint progressively increases, which in turn degrades the performance of CSI estimation. It is apparent from Fig. \ref{fig:DA_BL_iteration}(c) that the increased NMSE of the channel estimate at ultra-high bandwidths in the range $B = 100$ GHz in a dual-wideband THz system can be offset by augmenting the number of pilot blocks $M$. Moreover, the proposed PA-BL technique is seen to yield robust performance over a wide bandwidth range from $20$ GHz to $100$ GHz, even without any compensation for the dual-wideband effect, which is convenient for practical implementation. The proposed PA-BL framework is able to estimate the $N_R \in \left\{32,64\right\}$ $\times$ $N_T \in \left\{32,64\right\}$ dual-wideband THz MIMO channel using $M \in \left\{30, 50, 70, 90\right\}$ pilot blocks and $N_{RF} \in \left\{6,16\right\}$, with corresponding $MN_{RF}$ values of $\left\{180, 300, 420, 540\right\}$ for $N=6$ and $\left\{480, 800, 1120, 1440\right\}$ for $N=16$. Note that all these values satisfy $MN_{RF} \ll N_TN_R,$ where $N_TN_R = 1024$ for System-I and $N_TN_R = 4096$ for System-II. This leads to an under-determined system in Equation \eqref{vec_pilot_beam}, wherein the conventional LS/MMSE estimators cannot be applied for CSI estimation. In fact, the LS and MMSE require $M \approx 130$ pilot blocks for System-I and $M \approx 250$ pilot blocks for System-II, which increases the pilot overhead and reduces the spectral efficiency. This clearly illustrates the unique advantage of the proposed novel scheme for THz CSI estimation over the orthodox estimation paradigms. Fig. \ref{fig:DA_BL_iteration}(d) depicts the variation of the normalized array gain with respect to the number of antenna elements $N_R$. As $N_R$ increases, the spatial wideband effect in the THz system exhibits a significant impact due to the notable delay between the first and last antenna elements, leading to a decrease in the array gain. Consequently, it can be inferred that as $N_R \rightarrow \infty$, $G(\theta, f_k) \rightarrow 0$. Furthermore, Fig. \ref{sys1}(c) illustrates the variation of the normalized array gain with respect to the bandwidth $B$. A large bandwidth induces a significant shift in the relative subcarrier frequency $\delta_k = \frac{f_k}{f_c}$, leading to a significant degradation in normalized array gain. The results for the DA-BL framework that can lead to a further improvement in the performance of the CSI estimation are presented next.
\vspace{-5pt}
\subsection{DA-BL wideband THz hybrid MIMO channel estimation}
\begin{figure*}
	\centering
	\subfloat[]{\includegraphics[scale=0.42]{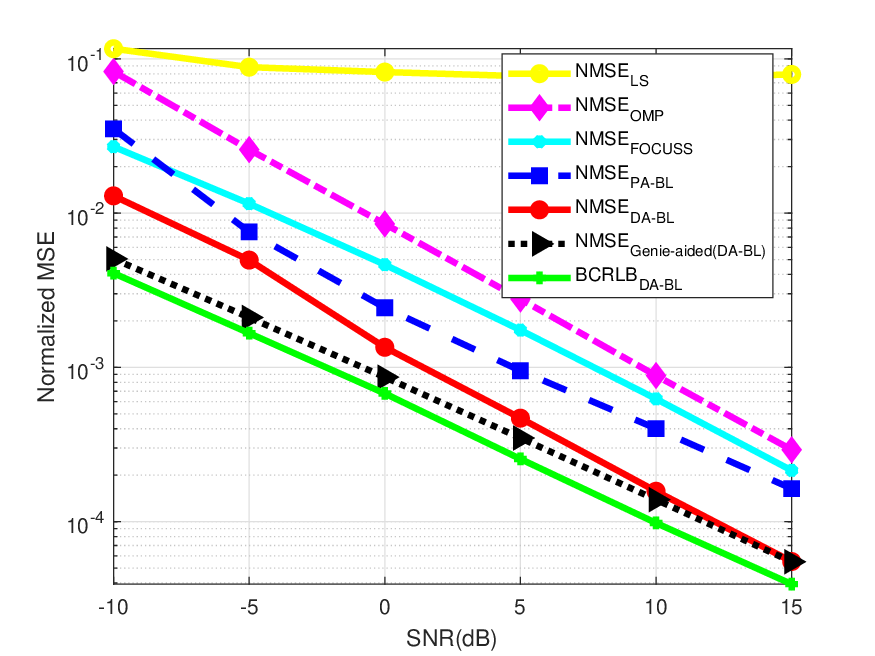}}
	\hfil
	\hspace{-10pt}\subfloat[]{\includegraphics[scale=0.42]{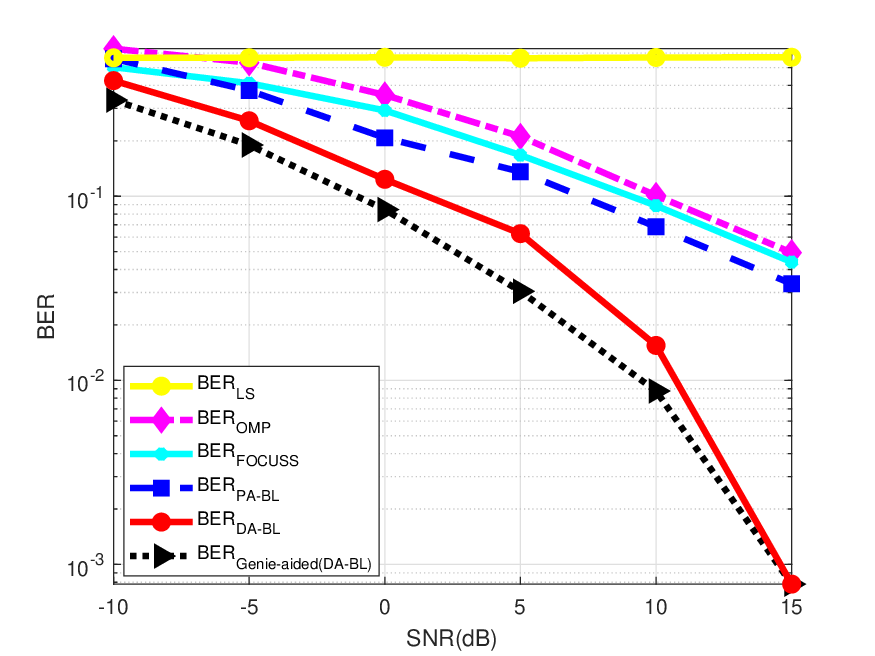}}
        \hfil
	\hspace{-10pt}\subfloat[]{\includegraphics[scale=0.42]{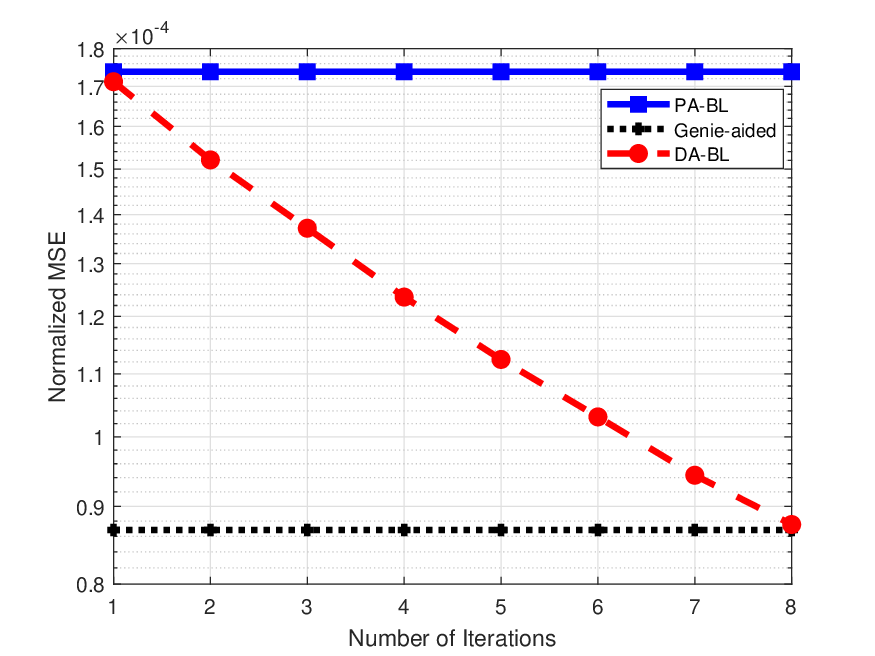}}
	\hfil
 \vspace{-4pt}
	  \caption{ $ \left(a\right) $ NMSE versus SNR comparison of the proposed PA-BL and DA-BL techniques with the Genie-aided estimator, LS, OMP, FOCUSS and the BCRLB, $ \left(b\right) $ BER versus SNR comparison of the proposed PA-BL and DA-BL techniques with the Genie-aided estimator for System-II. $ \left(c\right) $ NMSE of the DA-BL scheme with increasing number of iterations of DA-BL framework.}
	\label{sys2}
\end{figure*}

Fig. \ref{sys1} and \ref{sys2} demonstrate the performance improvement of the proposed PA-BL and DA-BL based techniques over traditional LS and other sparse estimation techniques such as OMP \cite{kulkarni2017low} and FOCUSS \cite{cotter2005sparse} along with a hypothetical scenario, where the true AoAs and AoDs are known, i.e, a Genie-aided transceiver for THz hybrid MIMO systems with parameter settings corresponding to System-I and System-II, as shown in Table II of the paper. Using the concatenated model for all the $M$ blocks as discussed in the context of equation \eqref{conc_data_pilot}, the hybrid TPC $\mathbf{F}_{RF}$ and hybrid RC $\mathbf{W}_{RF}$ are chosen to be block-independent, and are set as described in equation \eqref{frf_wrf}. As discussed, the performance of the traditional LS-based scheme is inferior, because the technique does not leverage the sparsity occurring in the THz MIMO channel. The performance of the OMP algorithm is adversely affected by its sensitivity to the selection of the stopping criteria and the choice of the dictionary matrix. Furthermore, FOCUSS suffers from convergence-related problems and it is highly sensitive to the regularization parameter. On the other hand, the PA-BL framework leverages sparsity through EM-based hyperparameter estimation. Moreover, it eliminates the need for tuning or regularization parameters and it is robust to the choice of dictionary matrix. However, the exclusive reliance on pilot symbols makes it challenging to attain a precise CSI estimate. Therefore, the proposed DA-BL demonstrates a superior performance compared to other sparse estimation techniques due to its ability to utilize the data symbols coupled with a minimal pilot overhead. Furthermore the DA-BL framework beneficially leverages the data symbols in addition to the pilot subcarriers to improve the channel estimate by iteratively maximizing the joint likelihood, as discussed in Section VI. The number of data vectors $N_d$ that contain the information symbols for the DA-BL framework of System I and II is set as $N_d = \left\{100,200\right\},$ respectively. The number of Monte Carlo iterations for the DA-BL framework is considered to be $DL_{max} = 20$ with the threshold of $\epsilon = 1$ as discussed in Algorithm 2. Additionally, it is clear from the figure that the performance of the proposed DA-BL framework approaches that of the Genie-aided system in the high SNR range, which evidences the efficiency of the proposed scheme. To benchmark the performance, the BCRLB is also plotted, as determined in Section V-A. The accuracy of data detection at the receiver using the MMSE receiver for the proposed PA-BL, DA-BL techniques and Genie-aided transceiver has been demonstrated by plotting the BER in Fig. \ref{sys1}(b) and similarly, for System-II, in Fig. \ref{sys2} (b). From \ref{sys1}(b) one can further observe that there is performance improvement of approximately $4$ dB is attained for the DA-BL framework over the PA-BL technique across the SNR range of $-10$ dB to $15$ dB.
For both the settings, the DA-BL-based receiver can be seen to yield a markedly improved performance in comparison to its PA-BL counterpart. Furthermore, the performance of the former, when System-II is considered, is also extremely close to that of a Genie-aided solution, and almost overlaps in the high-SNR regime, as it is clearly seen in Fig. \ref{sys2}(a). Fig. \ref{sys2}(c) illustrates the impact of the number of data-aided iterations on NMSE performance for SNR set as $10$ dB. Importantly, the performance of the PA-BL framework and Genie-aided estimator are insensitive to the number of data decoding iterations and are thus constant in this figure. One can also observe that the NMSE of the proposed DA-BL scheme achieves that of the Genie-aided receiver for as few as $8$ DA-BL iterations. This is of considerable importance, as the DA-BL framework achieves this without the need for prior knowledge of the true AoA/ AoD. This underscores its suitability for practical implementation in THz systems, where prior information is typically unavailable. Another noteworthy aspect is that, similar to the PA-BL receiver considered previously, for the setting with $M=30$ pilot blocks and $N_d = 100$ data vectors, the relation $(M+N_d)N_{RF} = 780 \ll N_TN_R = 1024$ once again holds true for the DA-BL receiver in System-I, which results in an under-determined estimation model as per Equation \eqref{vec_pilot_beam}. Thus, this clearly shows that even in such a scenario, wherein the conventional LS and MMSE estimators are ineffective, the dual-wideband channel can be estimated precisely, lower complexity and a notably smaller training overhead, employing the proposed Data-Aided framework. Moreover, the Bayesian learning algorithm has demonstrated superior performance compared to other sparse recovery techniques in scenarios where the spatial wideband effect is negligible \cite{srivastava2022hybrid}, \cite{srivastava2021data}. However, when the spatial wideband effect is significant, both the CSI estimation accuracy and the resultant BER performance degrade. This deterioration is attributed to the subcarrier-dependent nature of the transmit and receive dictionary matrices i.e $\mathbf{A}_T\left(\Theta_T, f_k\right)$ and $\mathbf{A}_R\left(\Theta_R, f_k\right)$ respectively. To address this issue, we seek a sparse recovery technique, which is robust to the choice of dictionary matrix and ensures convergence. In this context, the PA-BL algorithm emerges as an ideal choice. Additionally, to enhance the estimation accuracy and mitigate the impact of the spatial wideband effect, this work also proposes a data-aided Bayesian learning technique. This technique iteratively incorporates data symbol estimates along with the development of a robust ZF based data symbol detector. Notably, other sparse signal recovery techniques such as OMP, FOCUSS and LASSO perform poorly in the presence of the spatial-wideband effect. Furthermore, there is no systematic approach to formulate their data-aided extensions. Thus, the relative performance improvement of the Bayesian learning based schemes over OMP/ FOCUSS/ LASSO is significant in the presence of the spatial-wideband effect. In short, the excellent CSI estimation performance, coupled with lower overheads and robust data-detection, render the proposed frameworks eminently suitable for application in dual-wideband THz MIMO systems.
\section{Conclusions} 
New channel estimation paradigms considering angular domain sparsity in practical THz MIMO wideband channels have been conceived. The THz channel was modeled rigorously using parameters derived from the HITRAN database, considering the free-space losses in addition to the dual-wideband effect naturally occurring in such systems. An SC-FDE based I/O model was developed to beneficially transform the TD model to its equivalent FD representation. Subsequently, exploiting the sparsity of the dual-wideband THz channel, novel Bayesian learning approaches were proposed for CSI recovery. To begin with, the PA-BL methodology was conceived for channel estimation, that uses only pilot symbols, followed by the DA-BL approach for superior performance, that leverages the concatenated pilot and data outputs. By explicitly deriving the BCRLBs, we characterize the estimation performance of PA-BL and DA-BL frameworks. A comprehensive set of simulation results was presented that evidences the NMSE and BER performance benefits of the various schemes formulated in this paper. In particular, the NMSE of the DA-BL paradigm was seen to closely follow that of the Genie-aided transceiver and the BCRLB references at high SNRs, a feature that shows the advantages of practical deployment of the different algorithms devised.
\appendices
\section{ calculation of the matrix $\boldsymbol{\Xi}^{(j)}[k]$} \label{appealc}
The optimization cost function in \eqref{eq:decouple_max_4} can be simplified as 
\vspace{-15pt}
\begin{align}
& \mathbb{E} \left\{ \left\Vert \mathbf{Y}_{d}[k]-\mathbf{H}_{\textrm{eq}}[k]\mathbf{X}_{d}[k]\right\Vert_{F}^2 \right\}
= \notag \\ & \mathrm{Tr} \bigg\{ \mathbf{Y}_{d}^H[k] \mathbf{Y}_{d}[k] - \mathbf{Y}_{d}^H[k] \hat{\mathbf{H}}_{\textrm{eq}}^{(j)}[k]\mathbf{X}_{d}[k] - \notag \\ & \mathbf{X}_{d}^H[k] \left( \hat{\mathbf{H}}_{\textrm{eq}}^{(j)}[k]\right)^H \mathbf{Y}_{d}[k]- \mathbf{X}_{d}^H[k] \mathbb{E} \left\{ \mathbf{H}_{\textrm{eq}}^H[k] \mathbf{H}_{\textrm{eq}}[k]\right\} \mathbf{X}_{d}[k]  \bigg\}, 
\end{align}
where $\hat{\mathbf{H}}_{\textrm{eq}}^{(j)}[k]$ is given by \eqref{eq_ch_info}. Also the quantity $\mathbb{E} \left\{ \mathbf{H}_{\textrm{eq}}^H[k] \mathbf{H}_{\textrm{eq}}[k]\right\}$ is expressed as 
\vspace{-5pt}
\begin{align}
    \mathbb{E} \left\{ \mathbf{H}_{\textrm{eq}}^H[k] \mathbf{H}_{\textrm{eq}}[k]\right\} = \left(\hat{\mathbf{H}}_{\textrm{eq}}^{(j)}[k]\right)^H \hat{\mathbf{H}}_{\textrm{eq}}^{(j)}[k] + \boldsymbol{\Xi}^{(j)}[k].
\end{align}
The procedure for computation of the matrix $\boldsymbol{\Xi}^{(j)}[k]$ is described next. Let $\mathbf{h}_{\textrm{eq}}[k] \in \mathbb{C}^{N_{RF}^{2} \times 1}$ be the vectorized form of the equivalent channel matrix given by $\mathbf{h}_{\textrm{eq}}[k] =  \mathrm{vec} \left(\mathbf{H}_{\textrm{eq}}[k]\right)$ and its estimate $\hat{\mathbf{h}}_{\textrm{eq}}^{(j)}[k]$ be defined as $\hat{\mathbf{h}}_{\textrm{eq}}^{(j)}[k] = \mathrm{vec}\left( \hat{\mathbf{H}}_{\textrm{eq}}^{(j)}[k]\right)$. Its error covariance matrix $\boldsymbol\Sigma_{\textrm{eq}}^{(j)}[k] \in \mathbb{C}^{N_{RF}^{2} \times N_{RF}^{2}}$ can be derived by substituting the value of $\hat{\mathbf{H}}_{\textrm{eq}}^{(j)}[k]$ from \eqref{eq_ch_info} given
\vspace{-3pt}
\begin{align}
    \boldsymbol\Sigma_{\textrm{eq}}^{(j)}[k] = \widetilde{\boldsymbol{\Phi}}[k] \boldsymbol\Sigma^{(j)}[k] \widetilde{\boldsymbol{\Phi}}^H[k], \label{sigma_eq_m_j_k}
\end{align}
where $\widetilde{\boldsymbol{\Phi}}[k] = \left[ \mathbf{F}_{{RF}}^T \otimes \mathbf{W}_{{RF}}^H \right] \boldsymbol\Psi[k] \in \mathbb{C}^{N_{RF}^2\times G_R G_T}$. Subsequently, the $(a,b)$th element of the matrix $\boldsymbol{\Xi}^{(j)}[k]$ is defined
\vspace{-10pt}
\begin{align}
    \boldsymbol{\Xi}^{(j)}[k] (a,b) = & \mathrm{Tr}\big[ \boldsymbol\Sigma_{\textrm{eq}}^{(j)}[k] (\tilde{a}-N_{RF}+1 : \tilde{a}, \notag \\ &
    \tilde{b}-N_{RF}+1 : \tilde{b}) \big],
\end{align}
where $\tilde{a}  = a N_{RF}$ and $\tilde{b}  = b N_{RF}$.
\section{bayesian cram{\'e}r-rao lower bound for data-aided learning}
Let $\mathbf{J}_{\mathrm{DA}}[k] \in \mathbb{C}^{G_RG_T \times G_RG_T}$ denote the combined Bayesian FIM, which can be evaluated as
\vspace{-5pt}
\begin{align}
    \mathbf{J}_{\mathrm{DA}}[k] = \mathbf{J}_{d,\mathrm{DA}}[k] + \mathbf{J}_{b,\mathrm{DA}}[k],
\end{align}
where $\mathbf{J}_{d,\mathrm{DA}}[k] \in \mathbb{C}^{G_RG_T \times G_RG_T}$ denotes the FIM associated with the data output, while $\mathbf{J}_{b,\mathrm{DA}}[k]$ represents the FIM corresponding to the prior information of the beamspace CSI $\mathbf{h}_b[k]$. The above quantities can be computed as
\begin{align}
    \mathbf{J}_{d,\mathrm{DA}}[k] = -\mathbb{E}_{\mathbf{y}[k],\mathbf{h}_b[k]}\left\{\frac{\partial^2\mathcal{L}(\mathbf{y}[k]|\mathbf{h}_b[k];\mathbf{X}_d[k])}{\partial\mathbf{h}_b[k]\partial\mathbf{h}^H_b[k]}\right\} \notag
    \end{align}
    \vspace{-4pt}
    \begin{align}
    \mathbf{J}_{b,\mathrm{DA}}[k] = -\mathbb{E}_{\mathbf{h}_b[k]}\left\{\frac{\partial^2\mathcal{L}(\mathbf{h}_b[k];\mathbf{\Gamma}_k)}{\partial\mathbf{h}_b[k]\partial\mathbf{h}^H_b[k]}\right\}.
\end{align}
Using the conditional PDF as obtained in (67), the FIM $\mathbf{J}_{d,\mathrm{DA}}[k]$ can be obtained as 
\vspace{-5pt}
\begin{align}
    \mathbf{J}_{d,\mathrm{DA}}[k] = \tilde{\mathbf{\Psi}}[k]\mathbf{R}^{-1}_v\tilde{\mathbf{\Psi}}[k]. \label{final_FIM_data}
\end{align}
Similarly, as obtained in the PA-BL, the FIM associated with $\mathbf{h}_b[k]$ is obtained using the prior PDF from equation (68), and the corresponding expression for $\mathbf{J}_b[k]$ can be obtained as
\vspace{-5pt}
\begin{align}
    \mathbf{J}_{b,\mathrm{DA}}[k] = \hat{\mathbf{\Gamma}}_k^{-1}. \label{final_FIM_beam}
\end{align}
Using the expressions in \eqref{final_FIM_data} and \eqref{final_FIM_beam}, the Bayesian FIM can be represented as
\vspace{-5pt}
\begin{align}
\mathbf{J}_{\mathrm{DA}}[\mathit{k}] = \tilde{\mathbf{\Psi}}^{\mathit{H}}[\mathit{k}]\mathbf{R}_{\mathit{v}}^{-1}\tilde{\mathbf{\Psi}}[\mathit{k}] + \hat{\mathbf{\Gamma}}_{\mathit{k}}^{-1}. \label{FIM_sol}
\end{align}
Thus, the lower bound on the MSE of the estimate $\hat{\mathbf{h}}_{\mathit{b}}[\mathit{k}]$, can be obtained as
\vspace{-5pt}
\begin{align}
\begin{split}
\textrm{MSE}(\hat{\mathbf{h}}_{\mathit{b}}[\mathit{k}]) & = \mathbb{E}\left\{\parallel\hat{\mathbf{h}}_{\mathit{b}}[\mathit{k}]-\mathbf{h}_{\mathit{b}}[\mathit{k}]\parallel^{2}\right\}\geq \textrm{Tr}\left\{\mathbf{J}^{-1}_{\mathrm{DA}}[\mathit{k}]\right\} \\
& = \textrm{Tr}\left\{\left(\tilde{\mathbf{\Psi}}^{\mathit{H}}[\mathit{k}]\mathbf{R}_{\mathit{v}}^{-1}\tilde{\mathbf{\Psi}}[\mathit{k}] + \hat{\mathbf{\Gamma}}_{\mathit{k}}^{-1}\right)^{-1}\right\}. \label{mean_estimate}
\end{split}
\end{align}
Exploiting the relationship between the beamspace representation of the THz MIMO channel and its vectorized version, the BCRLB for the estimated CSI is finally given by
\vspace{-5pt}
\begin{align}
\textrm{MSE}(\hat{\mathbf{H}}[\mathit{k}]) \geq \textrm{Tr}\left\{\mathbf{\Psi}[\mathit{k}]\mathbf{J}^{-1}_{\mathrm{DA}}[\mathit{k}]\mathbf{\Psi}^{\mathit{H}}[\mathit{k}]\right\}. \label{bclb}
\end{align}
\bibliographystyle{IEEEtran}
\bibliography{main_1}
\begin{IEEEbiography}[{\includegraphics[width=1in,height=1.25in,clip,keepaspectratio]{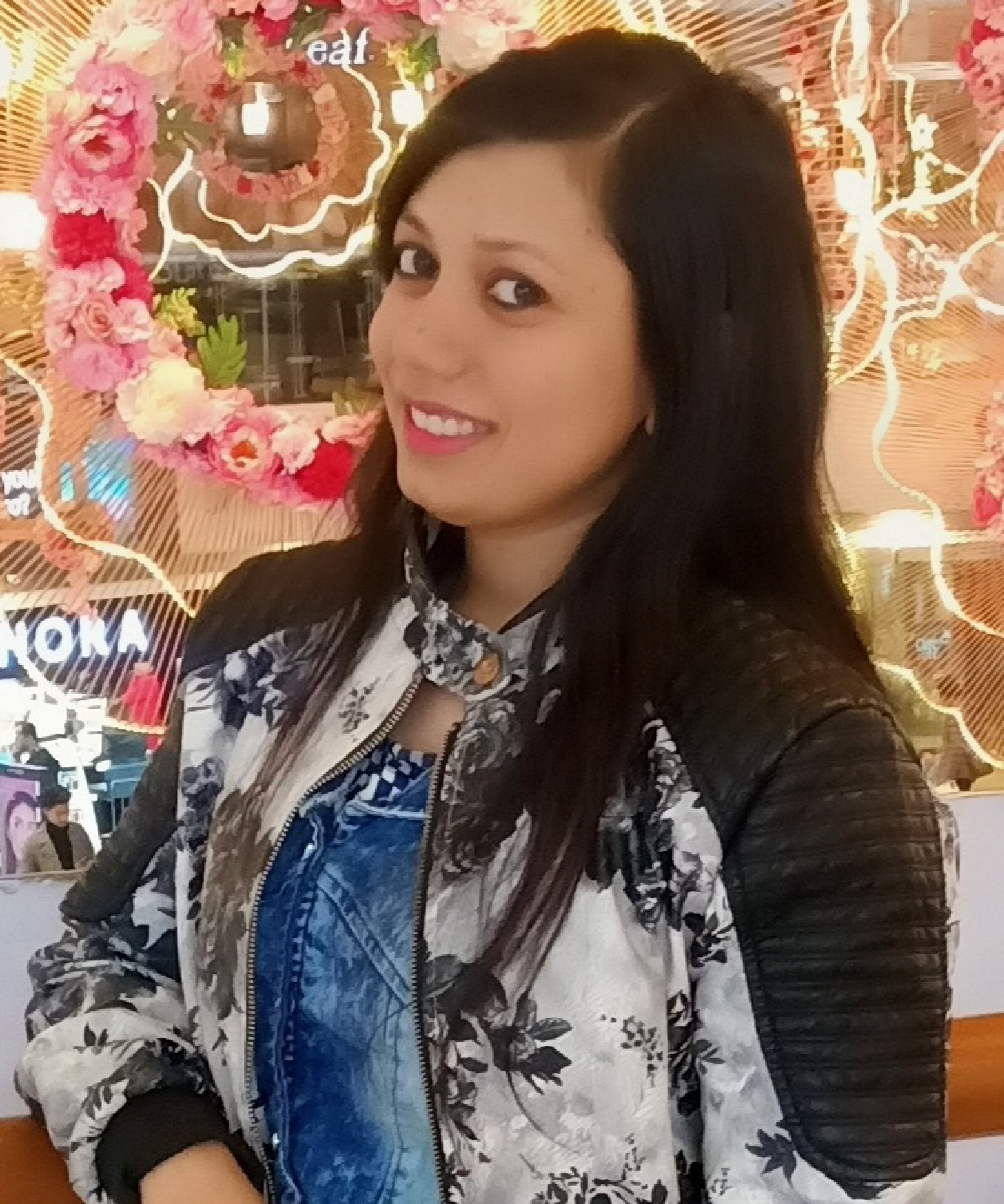}}]{Abhisha Garg}~(Graduate Student Member, IEEE) received the B.Tech degree in Electronics \& Communication from Graphic Era University, Dehradun, India, in $2016$, and the M.Tech degree in Signal Processing from Guru Gobind Singh Indraprastha University, New Delhi, India, in $2020$. She is currently working toward the Ph.D. degree with the Department of Electrical Engineering, Indian Institute of Technology Kanpur, Kanpur, India. Her area of research includes THz communications, sparse signal processing, machine learning and its applications. She was awarded with the Gold medal from Graphic Era University, Dehradun, India for academic excellence. She was also awarded the Young Scientist award from International Science Community Association (ISCA), in $2017$. She was a recipient of IEEE Signal Processing Society (SPS) scholarship, $2023$. She was also a finalist for the Qualcomm Innovation Fellowship (QIF), $2023$.
\end{IEEEbiography}
\begin{IEEEbiography}[{\includegraphics[width=1in,height=1.25in,clip,keepaspectratio]{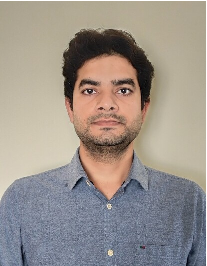}}]{Suraj Srivastava}~(Member, IEEE) received the M.Tech. degree in Electronics and Communication Engineering from Indian Institute of Technology Roorkee, India, in $2012$, and Ph.D. degree in Electrical Engineering from Indian Institute of Technology Kanpur, India, in $2022$. From July $2012$ to November $2013$, he was employed as a Staff-I systems design engineer with Broadcom Research India Pvt. Ltd., Bangalore, and from November $2013$ to December $2015$, he was employed as a lead engineer with Samsung Research India, Bangalore where he worked on developing layer-$2$ of the $3$G UMTS/WCDMA/HSDPA modem. He worked as a senior lead engineer in Qualcomm India Pvt. Ltd., Bangalore, from October $2022$ to November $2023$. His research interests include applications of Sparse Signal Processing in 5G Wireless Systems, mmWave and Tera-Hertz Communication, Orthogonal Time-Frequency Space (OTFS), Joint Radar and Communication (RadCom), Optimization and Machine Learning. He was awarded Qualcomm Innovation Fellowship (QIF) in year $2018$ and $2022$ from Qualcomm. He was also awarded Outstanding Ph.D. Thesis and Outstanding Teaching Assistant awards from IIT Kanpur. He is currently an Assistant Professor at the Department of Electrical Engineering, Indian Institute of Technology Jodhpur.
\end{IEEEbiography}
\begin{IEEEbiography}[{\includegraphics[width=1in,height=1.25in,clip,keepaspectratio]{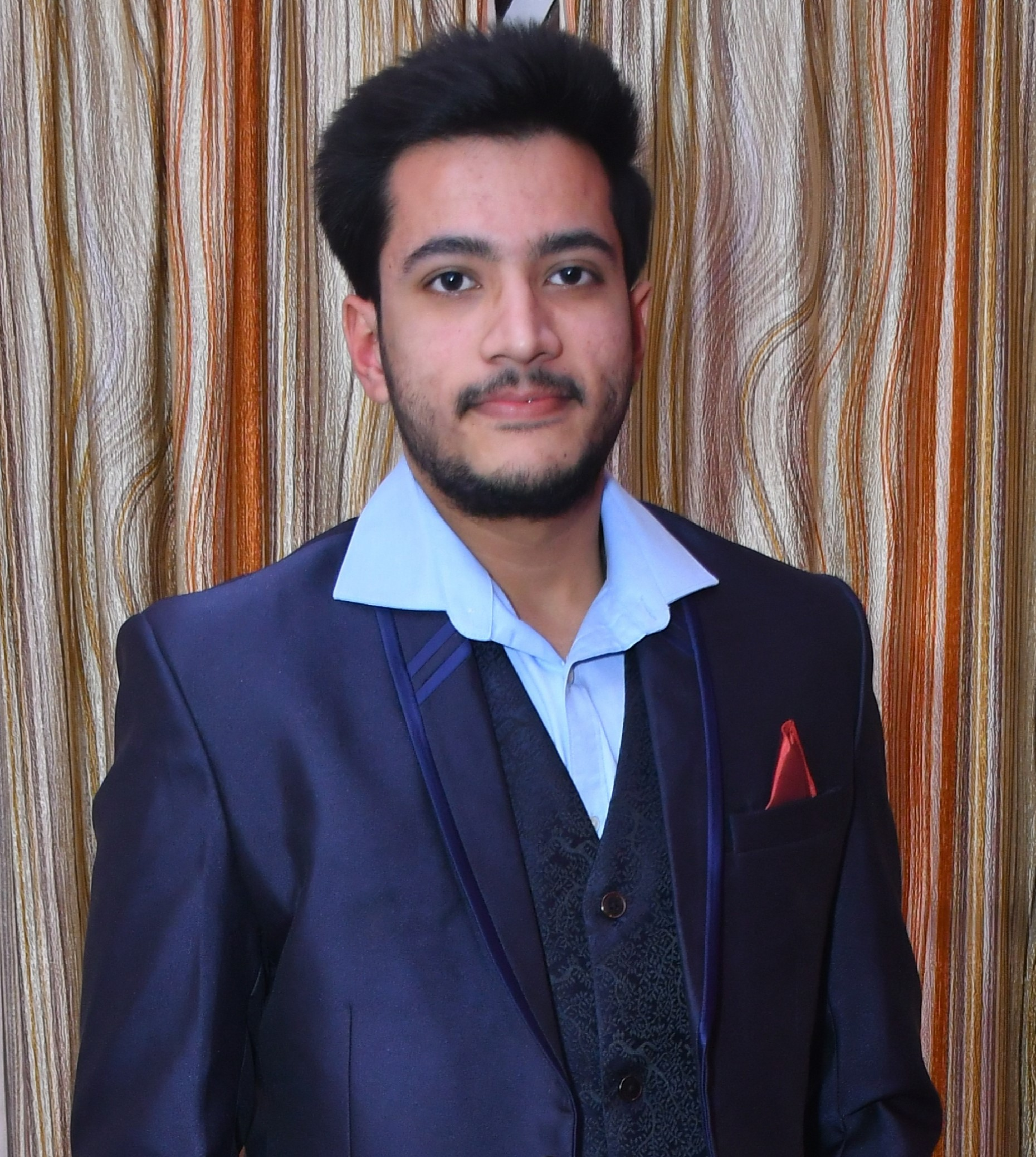}}]{Nimish Yadav} received the B.Tech. degree in Electronics Engineering from Kamla Nehru Institute of Technology, Sultanpur, India, in $2020$ and the M.Tech degree in Signal Processing and Communication from Department of Electrical Engineering, Indian Institute of Technology Kanpur, Kanpur, India, in $2022$. He is currently working with Qualcomm India Pvt. Ltd., Chennai, India, designated as a Hardware Systems Engineer in Wireless LANs. His research interests include sparse signal processing and THz communication. He was awarded with the Bronze medal from Dr. A.P.J. Abdul Kalam Technical University, Uttar Pradesh, India for academic excellence.
\end{IEEEbiography}
\begin{IEEEbiography}[{\includegraphics[width=1in,height=1.25in,clip,keepaspectratio]{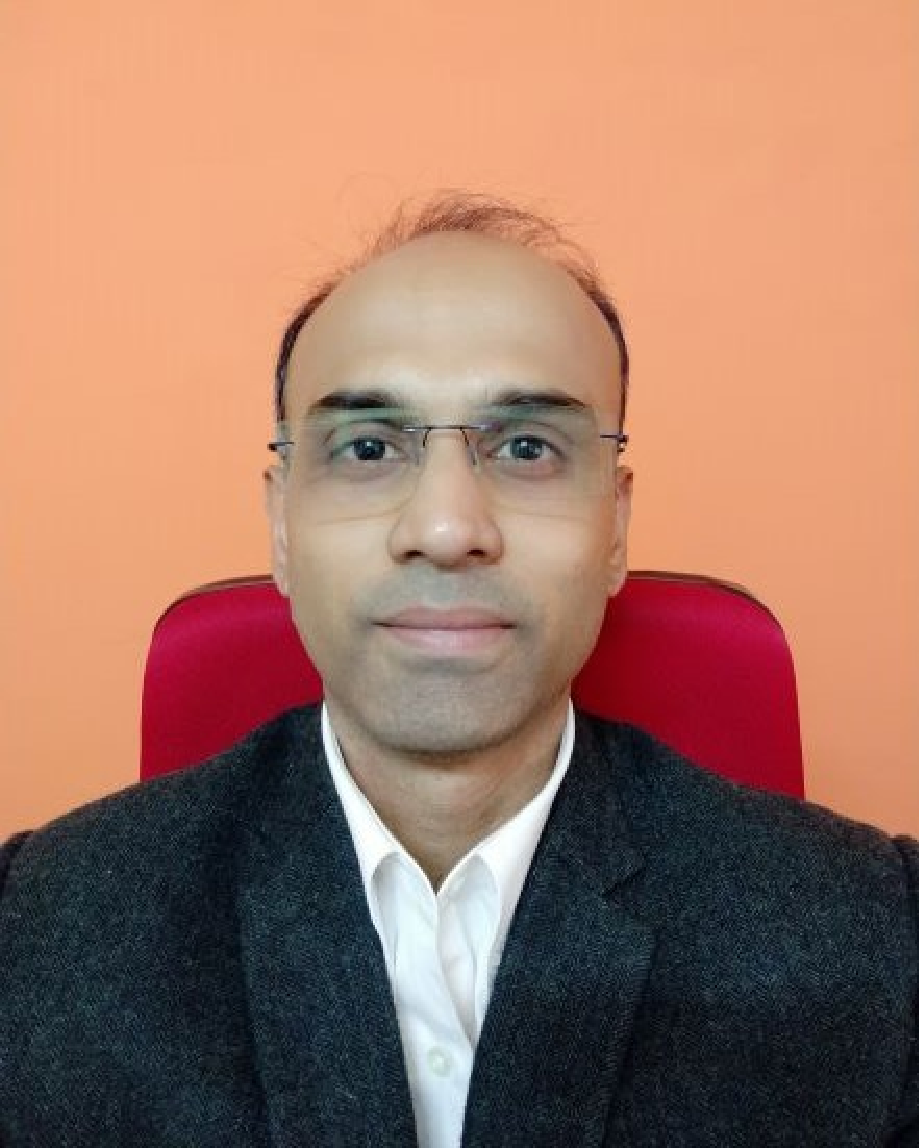}}]{Aditya K. Jagannatham}~(Senior Member, IEEE) received the bachelor’s degree from the Indian Institute of
Technology, Bombay, and the M.S. and Ph.D. degrees from the University of California at San Diego, San Diego, CA, USA. From April $2007$ to May $2009$, he was employed as a Senior Wireless Systems Engineer with Qualcomm Inc., San Diego, where he was a part of the Qualcomm CDMA Technologies (QCT) Division. He is currently a Professor with
the Department of Electrical Engineering, IIT Kanpur, where he also holds the Arun Kumar Chair Professorship. His research interests include next generation wireless cellular and WiFi networks, with a special emphasis on various $5$G technologies such as massive MIMO, mmWave MIMO, FBMC, NOMA, as well as emerging $6$G technologies such as OTFS, IRS, THz systems and VLC. He has been twice awarded the P. K. Kelkar Young Faculty Research Fellowship for excellence in research, received multiple Qualcomm Innovation Fellowships (QIF $2018, 2022$), the IIT Kanpur Excellence in Teaching Award, the CAL(IT)$2$ Fellowship at the University of California at San Diego, the Upendra Patel Achievement Award at Qualcomm San Diego and the Qualcomm $6$G UR India gift.
\end{IEEEbiography}
\begin{IEEEbiography}[{\includegraphics[width=1in,height=1.25in,clip,keepaspectratio]{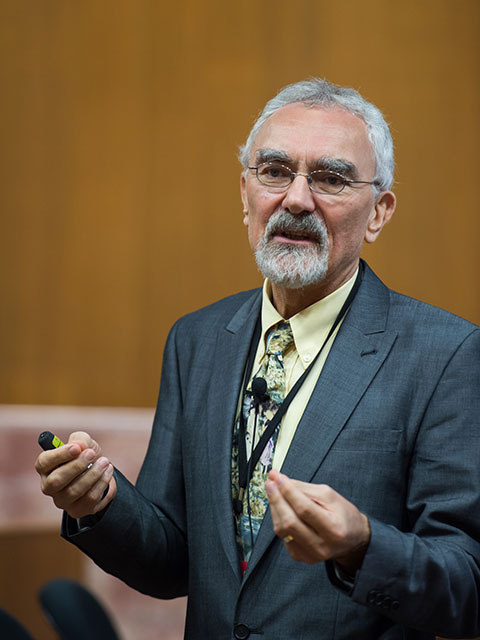}}]{Lajos Hanzo}~(Life Fellow, IEEE) received Honorary Doctorates from the Technical University of Budapest ($2009$) and Edinburgh University ($2015$). He is a Foreign Member of the Hungarian Science-Academy, Fellow of the Royal Academy of Engineering (FREng), of the IET, of EURASIP and holds the IEEE Eric Sumner Technical Field Award. For further details please see \url{http://www-mobile.ecs.soton.ac.uk/}, \url{https://en.wikipedia.org/wiki/Lajos_Hanzo}.
\end{IEEEbiography}
\end{document}